\documentclass[fleqn]{SCYE}
\usepackage{longtable}
\usepackage{verbatim}
\usepackage{multirow}
\usepackage[numbers,sort&compress]{natbib}

\begin{document}


\ArticleType{Article}
\Year{2020}

\title{Concept of the Solar Ring Mission: Overview}{Concept of the Solar Ring Mission: Overview}

\author[1,2,]{Yuming WANG}{{ymwang@ustc.edu.cn}}%
\author[3]{Haisheng JI}{}
\author[4,5]{Yamin WANG}{}%
\author[6]{Lidong XIA}{}
\author[1,2]{Chenglong SHEN}{}
\author[1,2]{\\Jingnan GUO}{}
\author[1,2]{Quanhao ZHANG}{}
\author[6]{Zhenghua HUANG}{}
\author[1,2]{Kai LIU}{}
\author[1,2]{Xiaolei LI}{}
\author[1,2]{Rui LIU}{}
\author[7]{\\Jingxiu WANG}{}
\author[1,2]{Shui WANG}{}

\AuthorMark{Wang Y}

\AuthorCitation{Wang, Y., H. Ji, Y. Wang, L. Xia, C. Shen, J. Guo, Q. Zhang, K. Liu, X. Li, R. Liu, J. Wang, and S. Wang}

\address[1]{CAS Key Laboratory of Geospace Environment, School of Earth and Space Sciences, University of Science and Technology of China,\\ Hefei 230026, China}
\address[2]{CAS Center for Excellence in Comparative Planetology, USTC, Hefei 230026, China}
\address[3]{Purple Mountain Observatory, Chinese Academy of Sciences, Nanjing 210034, China}
\address[4]{CAS Key Lab of Microsatellite, Innovation Academy for Microsatellites, Chinese Academy of Sciences, Shanghai 201203, China}
\address[5]{Shanghai Engineering Center for Microsatellites, Shanghai 201203, China}
\address[6]{Shandong Provincial Key Laboratory of Optical Astronomy and Solar-Terrestrial Environment, Institute of Space Sciences,\\ Shandong University, Weihai 264209, China}
\address[7]{Univeristy of Chinese Academy of Sciences, Beijing 100049, China}


\abstract{The concept of the Solar Ring mission was gradually formed from L5/L4 mission concept, and the proposal of its pre-phase
study was funded by the National Natural Science Foundation of China in November 2018 and then by the
Strategic Priority Program of Chinese Academy of Sciences in space sciences in May 2019. Solar Ring mission will be the
first attempt to routinely monitor and study the Sun and inner heliosphere from a full $360$-degree
perspective in the ecliptic plane. The current preliminary design of the Solar Ring mission is to deploy six spacecraft, grouped in three
pairs, on a sub-AU orbit around the Sun. The two spacecraft in each group
are separated by about $30^\circ$ and every two groups by about $120^\circ$. This configuration with necessary
science payloads will allow us to
establish three unprecedented capabilities: (1) determine the photospheric vector magnetic
field with unambiguity, (2) provide $360$-degree maps of the Sun and the inner heliosphere routinely, and (3) resolve the solar wind structures
at multiple scales and multiple longitudes. With these capabilities, the Solar Ring mission aims to address
the origin of solar cycle, the origin of solar eruptions, the origin of
solar wind structures and the origin of severe space weather events. The successful accomplishment of the mission
will advance our understanding of the star and the space environment that hold our life and enhance our capability of
expanding the next new territory of human.}



\maketitle


\begin{multicols}{2}
\section{Introduction}\label{sec:intro}

As the development of technology, the boundary of human explorations was and is being constantly expanded,
from continents, oceans, sky to space and even other planets. In near future, we believe,
the deep space and other terrestrial
planets, like Mars, will be the next new territory of human.
\Authorfootnote \noindent
During the expansion of human activities into the deep space, we have to understand the
Sun, interplanetary space and the space environments of the Earth and other planets.

Our Sun, the nearest star in the universe, primarily
controls the electromagnetic radiation
and particle radiation
environments of the (inter)planetary spaces in either short term through
various explosive activities or long term through
solar cycles or even longer periodic variations. All these short
and long term activities can be treated as the manifestations
and results of the changes of the magnetic field of the Sun.
The huge amount of magnetic energy in the solar corona, accumulated due to the plasma flows
in the lower solar atmosphere and/or even below photosphere, provides sufficient
free energy for violent solar eruptions, e.g., flares and coronal mass
ejections (CMEs). A typical solar eruption will release the energy of $10^{25}$ J in various
forms along with the mass of $10^{12}$ kg and the magnetic flux of $10^{15}$ Wb\citep{Hudson_etal_2006},
that will significantly disturb interplanetary space and may cause severe space weather events
in the time scale from minutes to days.

Solar steady outflows and eruptions, making up the solar wind, travel through interplanetary space
with a speed of more than hundreds of kilometers per second, and impact our planets. The in-situ
observations of the solar wind in the past four decades have shown that the changes in magnetic
field strength, plasma density, temperature and bulk velocity may exceed two orders of magnitude,
and the flux of energetic particles can be enhanced by four or even larger orders during an event.
Such large variations reflect the various levels of the energy and mass released from the Sun, and
could severely affect the satellites and astronauts in the space.
Thus, space weather forecasting has become an extremely important topic with
significant application values for high-tech systems, especially for the human activities
in the deep space.

The global and local dynamo processes make the solar activity gradually vary with multiple
periods, among which the most famous one is the quasi-11-year solar cycle. About every 11 years,
solar activity level increases from minimum to maximum and returns back to minimum accompanied
with the reversal of magnetic field polarities between the south and north poles. Aforementioned
solar eruptions are nearly ten times more frequent during solar maxima than during solar
minima\citep{Yashiro_etal_2004}. Solar interior structure and processes are one of the keys
to understand these periods\citep{Dikpati_Charbonneau_1999}. Besides, solar minima seemingly affect the space environment and
the Earth's system, including the human society and civilization, more lasting and profound\citep{Reid_1999,
Lean_Rind_1999} than solar maxima.
In the last solar cycle, we experienced a deep solar minimum, which is the deepest in the past
half of a century\citep{Nandy_etal_2011, Schrijver_etal_2011a, McComas_etal_2013}.
Will we experience another even deeper solar minimum and is this the
start of a new little ice age\citep{Feulner_Rahmstorf_2010, Schrijver_etal_2011a}? These questions have become the significant science
issues of solar physics, space physics, and earth sciences.

These current knowledge of our Sun and (inter)planetary space are particularly owing to continuous
space science missions in the past decades. From the view of Earth, we have the SOlar and Heliospheric
Observatory (SOHO)\citep{Domingo_etal_1995}, the Transition Region And Coronal Explorer (TRACE)\citep{Handy_etal_1999},
Yohkoh\citep{Ogawara_etal_1991}, the Solar Dynamic Observatory (SDO)\citep{Pesnell_etal_2012}, Hinode\citep{Kosugi_etal_2007}, etc. Since the
successful launch of the Solar TErrestrial RElations Observatory (STEREO) in 2006\citep{Kaiser_etal_2008}, human
for the first time watched the Sun and heliosphere simultaneously from two perspectives. In 2020,
human will be able to obtain unprecedented images of the Sun off the ecliptic plane with Solar
Orbiter (SolO)\citep{Muller_etal_2013}. Except these imaging-enabled missions, we also have many space missions
to sample local solar wind plasma, energetic particles and magnetic field, like the spacecraft
Wind\citep{Ogilvie_Parks_1996}, the Advanced Composition Explorer (ACE)\citep{Stone_etal_1998}, the Deep Space
Climate ObserVatoRy (DSCOVR)\citep{dscovr_2015} at L1 point of the Sun-Earth system, the Helios\citep{Winkler_1976}
in the inner heliosphere, and the Ulysses\citep{Wenzel_etal_1992} on a large elliptical polar orbit at about 5 AU.
The recently launched Parker Solar Probe (PSP) will eventually fly to a distance of $8.5$ solar radii from the Sun to
sample the solar corona\citep{Fox_etal_2016}. Besides, planetary science missions, e.g., the MErcury
Surface, Space ENvironment, GEochemistry, and Ranging (MESSENGER)\citep{Solomon_etal_2007}, the Venus Express\citep{Svedhem_etal_2007},
the Mars Express\citep{Schmidt_2003}, and the Mars Atmosphere and Volatile Evolution (MAVEN)\citep{Jakosky_etal_2015},
provide additional information of the space environment near the planets. More than 20-year data
from these great missions kept advancing our understanding on our Sun and (inter)planetary space.

However, we have not yet achieved the real-time observations of the full solar disk in 360 degrees,
which are essential to understand the whole evolution process of a sunspot, an active region (AR) and a
coronal hole (CH) from their birth to death, to infer the solar internal structure,
and to make long-term space weather forecasting possible.
We have not yet achieved the unambiguous observations of the photospheric vector magnetic fields,
which are the basis to understand all kinds of explosive phenomena on the Sun, and to realize how
the local and global dynamos work. We have not yet achieved the routine observations combining the
in-situ measurements and the remote panorama images of the solar wind and transients fully covering
the inner heliosphere, which is the only way to understand the evolution of the solar outflows and
eruptive structures, and to evaluate and forecast their space weather effects on our planets.
Now we propose a new space scientific mission, Solar Ring, to accomplish the above capabilities.

\section{Scientific Rationale and mission objectives}\label{sec:obj}

\begin{figure*}[t]
\centering
\includegraphics[width=0.48\hsize]{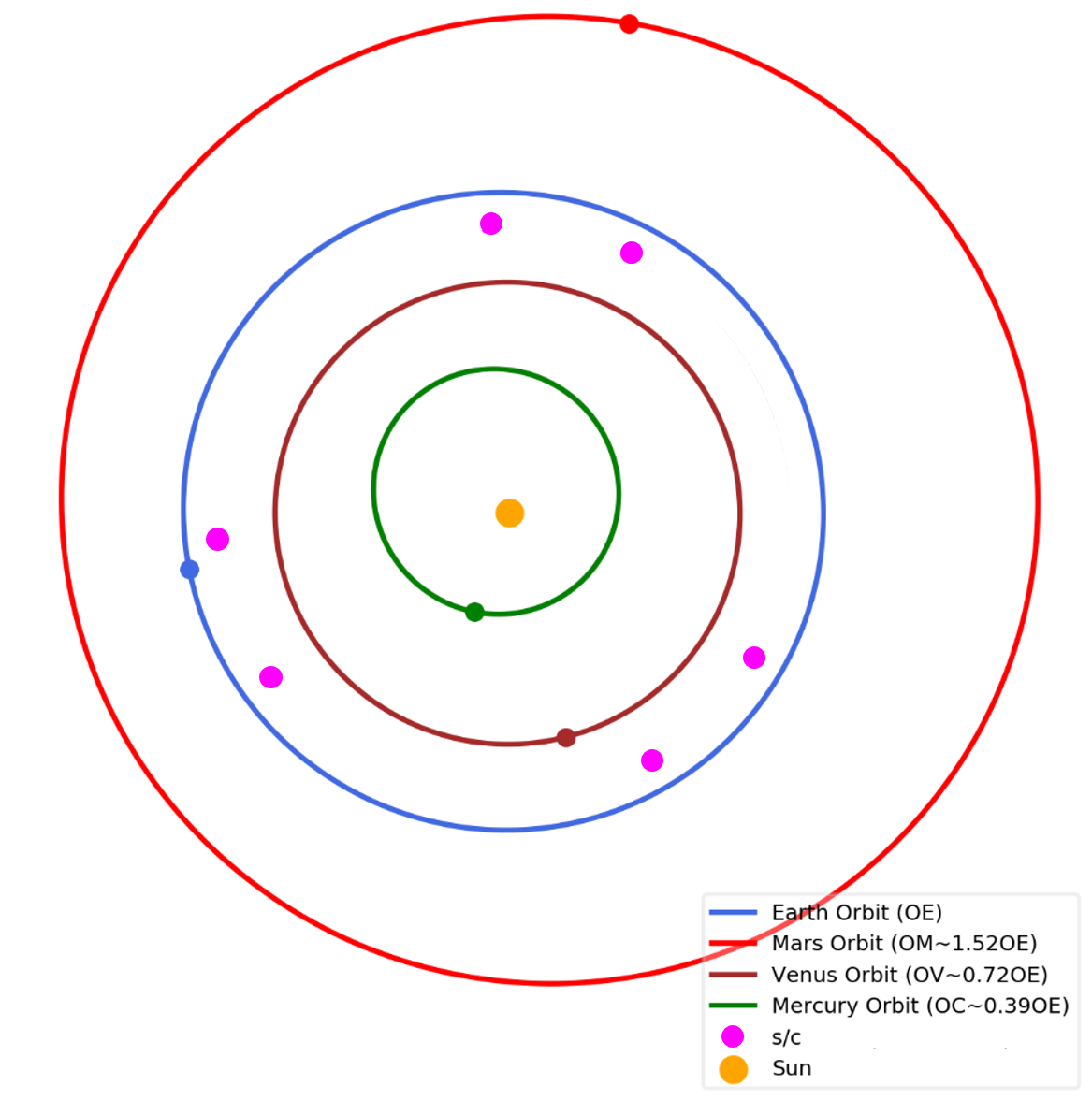}
\includegraphics[width=0.46\hsize]{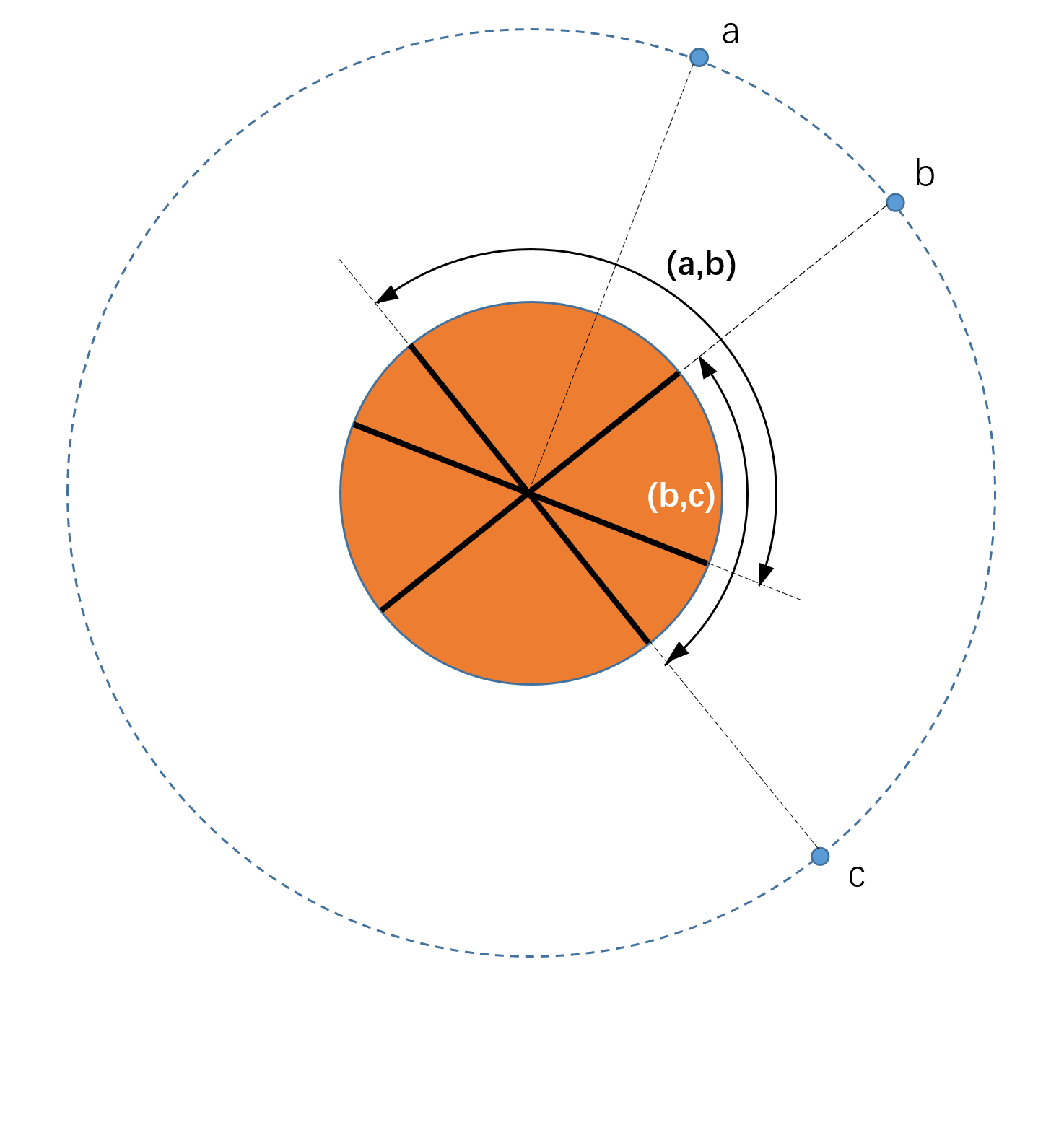}
\caption{{\it Left panel:} Schematic diagram of the Solar Ring mission. The six pink dots roughly denote the positions of the six spacecraft, which run on a sub-AU orbit. {\it Right panel:} A sketch map illustrates the stereoscopic angles of the spacecraft and their coverage in the longitude.}
\label{fg:sc_config}
\end{figure*}

The preliminary concept of the Solar Ring mission is to deploy 6 spacecraft circling around the Sun
at a sub-AU distance in the ecliptic plane as illustrated in Figure~\ref{fg:sc_config}a.
The idea was first from the L5/L4 mission concept, in which one spacecraft is suggested to operate at
L5 point, the upstream of the Earth, to monitor the space weather in advance, and one at L4
point, the downstream of the Earth, to get the effect of the space weather. The application value of
such a mission is obvious and important. To enrich and enhance its science merits,
we started to think about an upgraded one since the summer of 2017, and then gradually form the idea presented here,
of which the scientific goals of L5/L4 mission are included and extended.
Now the concept study has been funded by National Natural Science Foundation of China in the end of 2018
and by the Strategic Priority Program of Chinese Academy of Sciences in space sciences in early 2019.

The preliminarily designed orbit of Solar Ring, which is an elliptical orbit inside the Earth orbit (see
Sec.\ref{sec:design} and the companion paper\citep{WangY_etal_2020} for details), is a compromise between the
scientific goals and the cost of the launch and transfer of the spacecraft.
The spacecraft can self-drift to the desired position after being inserted into
the orbit just like STEREO, that requires less fuel and can carry more scientific payloads.
The six spacecraft are grouped in three pairs. In each pair, the two spacecraft are separated by about
$30^\circ$, and between the pairs, the separation angle is about $120^\circ$ (or about $90^\circ$
between the two closest spacecraft). This deployment not only
achieves the entire $360^\circ$ view of the Sun, but also provides different stereoscopic views with
the angle of about $30^\circ$, $90^\circ$, $120^\circ$ and $150^\circ$. By using this configuration, Solar
Ring mission will perform high resolution imaging of from photosphere to inner heliosphere and quasi-heliosynchronous
in-situ sampling of particles and fields. The mission will address the following four major scientific
themes:
\begin{itemize}
\item Origin of solar cycle 
\item Origin of solar eruptions
\item Origin of solar wind structures 
\item Origin of severe space weather events
\end{itemize}
through the three unprecedented capabilities listed below.

\paragraph{Measure photospheric vector magnetic fields with unambiguity\\}

Photospheric magnetic field is so far the only vector magnetic field in the space that can be remotely measured by
human. Hence, it is so far the only key to understanding our magnetized star.
The basic principle is that the spectral lines will split and get polarized in the presence of a magnetic
field due to Zeeman effect. Though the Sun's magnetic field has been measured for more than 110 years,
we still has not got accurate measurements of the photospheric vector magnetic fields without unambiguity. The main reason is
that the direction of the transversal component has so called
$180^\circ$ ambiguity\citep{Gary_Hagyard_1990}. Besides, the transversal component of the measured photospheric
magnetic field carries an uncertainty about one order higher in magnitude than the longitudinal (or line-of-sight) component.

\begin{figure}[H]
\centering
\includegraphics[width=\hsize]{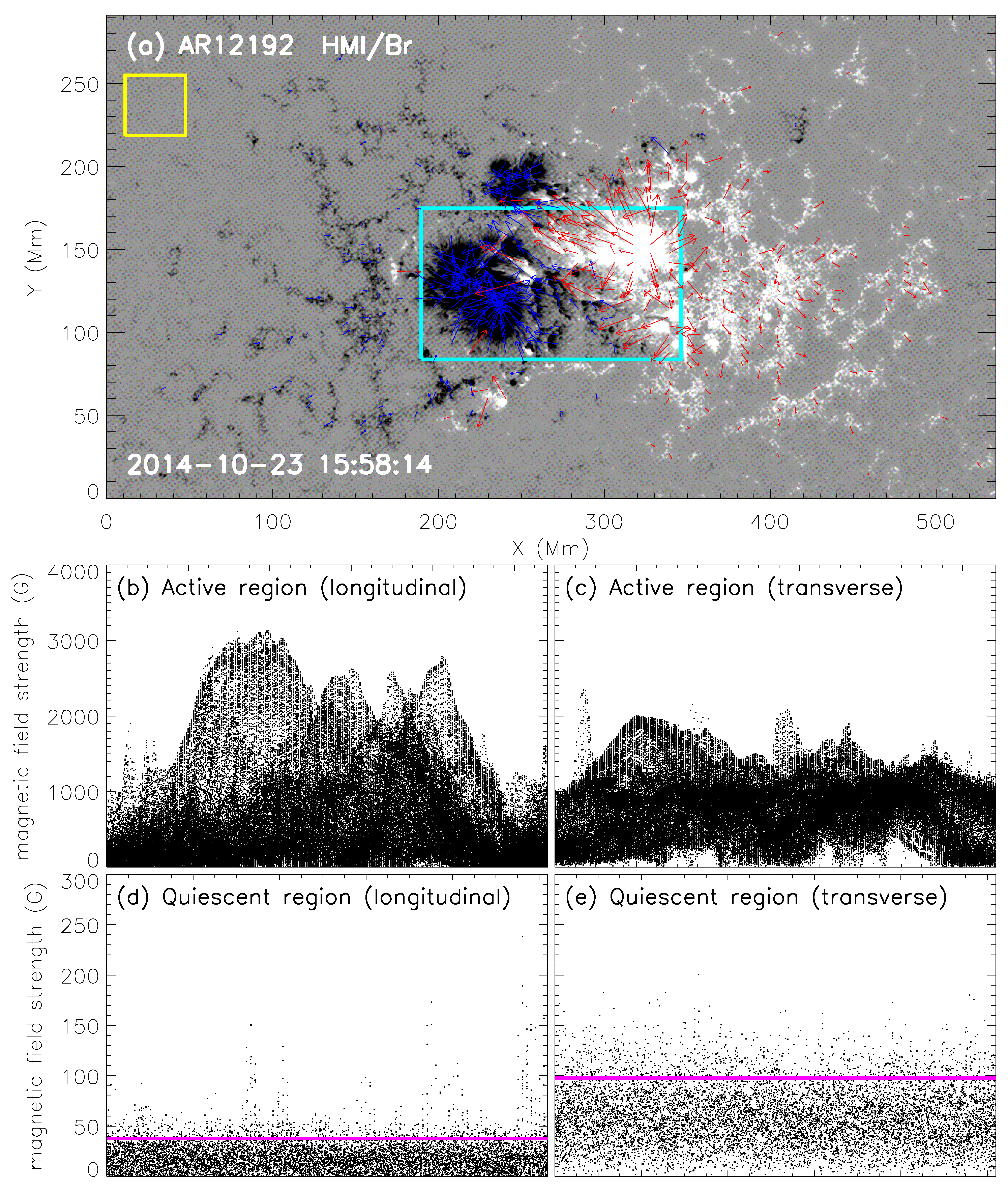}
\caption{{\it Panel (a):} The HMI vector magnetogram of the active region NOAA 12192. The gray color scales the longitudinal component
of the magnetic field, and the red/blue arrows denote the transversal component. {\it Panel (b) and (c):} Scatter plots of the longitudinal
and transversal components of the magnetic field in the central region marked by the cyan box in Panel (a). {\it Panel (d) and (e):}
The same scatter plots, but in the region near the edge marked by the yellow box in Panel (a). The horizontal axes in Panels (b)--(e)
mark the serial numbers of the data points in the data set, having no physical meaning. The magnetic field within the active region is
highly structured, but that in the quiescent region is more close to noise. The horizontal pink lines denote the 90-percentile of the
magnetic field strength, suggesting the noise level.}
\label{fg:hmi}
\end{figure}

The Helioseismic and Magnetic Imager (HMI)\citep{Schou_etal_2012} on board SDO measures the photospheric vector
magnetic fields on the full solar disk. For example, Figure~\ref{fg:hmi} shows the vector magnetogram of
AR 12192\citep{LiuL_etal_2016} and the scatter plots of the magnetic field strength in the AR and outside the AR, respectively.
It is clear that the magnetic field inside the AR is highly structured but that outside the AR is random, indicating
the noise. The horizontal lines in the bottom panels mark the 90-percentile of the magnetic field strength, which
could be treated as the uncertainty of the measurements. For the longitudinal component (Fig.\ref{fg:hmi}d),
the uncertainty is less than $40$ G, whereas that of the transversal component is about $100$ G. Since the
photospheric magnetic fields in quite Sun regions and CHs are typically of the order of ten
Gauss, the measured transversal component is only reliable and applicable in ARs, where
the magnetic field is strong enough. It should be aware that the observed magnetic field strength mentioned here
is actually the magnetic flux density, an average of the magnetic fields over a certain area depending on the spatial
resolution. Using high-resolution observations, magnetic fields as high as hundreds of Gauss are often
found in quiet Sun regions\citep{JinC_etal_2012}.

Above the photosphere, there are the chromosphere, transition region, corona and interplanetary space, where
more key processes of the solar eruptions, coronal heating and solar wind acceleration happen. However,
due to the low density, high temperature and highly dynamic atmosphere above the photosphere, the three dimensional (3D)
magnetic field has never been precisely measured. Most of the information of the coronal magnetic fields
come from the extrapolation of the photospheric magnetic fields\citep{Wiegelmann_Sakurai_2012}.
Non-linear force-free field
(NLFFF) extrapolation is a widely used approach to reveal the evolution of the coronal magnetic energy
before and after an eruption. All the force-free field extrapolations are model dependent.
As a most widely used NLFFF model developed by Wiegelmann\citep{Wiegelmann_2008}, for instance, it has been successfully applied to
study many solar eruptive events. A recent application of this model is to identify and study the solar
magnetic flux ropes during an eruption, but an interesting thing is that none of the identified
magnetic flux ropes stays above $10$ Mm from the solar surface, inconsistent with the frequently-found high
lying prominences/filaments. This is due to the treatment of energy minimization when processing photospheric
vector magnetograms, which is basically the limitation of the presence of the $180^\circ$ ambiguity in the
transversal component. Thus, such inaccurate measurements of photospheric vector magnetic fields largely
limit our understanding of the solar activities.

\begin{figure}[H]
\centering
\includegraphics[scale=0.25]{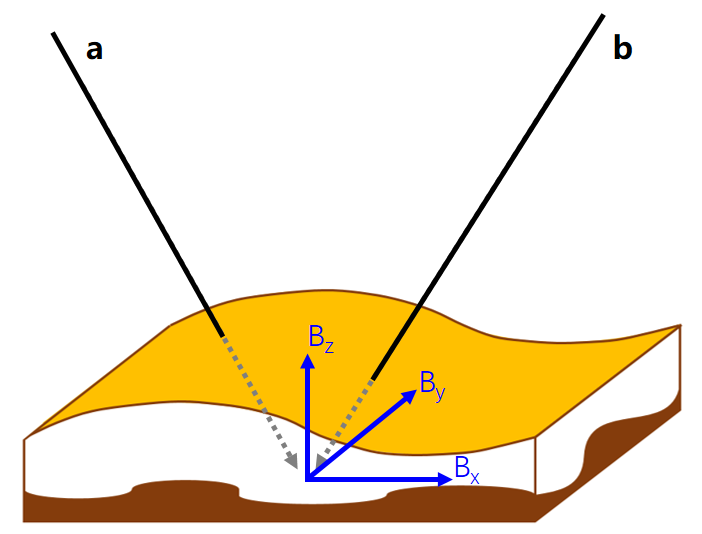}
\caption{A cartoon of the local photospheric region, illustrating some issues affecting the accuracy of the inversion
of magnetic field from dual perspectives.
The black arrows a and b indicate the observational paths. The photosphere is a non-uniform layer with different optical depths.}
\label{fg:bvec}
\end{figure}

For remote sensing, the only way to remove the $180^\circ$ ambiguity and increase the measurement accuracy is the multiple-perspective
observation. Three spacecraft forming a solid angle of a certain value (not too small and not too large) to observe circular
polarization will be the cleanest way to accurately map magnetic fields of the photosphere. However, since flying away from ecliptic
plane is technically difficult and very expensive, two spacecraft observing both circular and linear polarization will be
the feasible plan for us to remove the $180^\circ$ ambiguity (the detailed analysis is being prepared in the follow-up paper\citep{WangY_etal_2020a}).
Though STEREO spacecraft have been able to view the Sun from two perspectives off the Sun-Earth
line since 2006, unfortunately they did not carry a magnetic imager. Our mission will solve this issue by providing vector
magnetic observations from multiple perspectives as mentioned above. In principle, if there are vector magnetic
measurements from two viewing angles as illustrated in Figure~\ref{fg:bvec}, the magnetic field vector can be
inversed by using the formulae
\begin{eqnarray}
\left\{\begin{array}{l}
\mathbb M^a\vec{B}=B^a_L\hat{r}^a\pm\vec B^a_T\\
\mathbb M^b\vec{B}=B^b_L\hat{r}^b\pm\vec B^b_T
\end{array}\right.
\end{eqnarray}
in which $\mathbb M$ is the transform matrix, $B_L$ and $\vec B_T$ are the longitudinal and transversal components
of the measured magnetic field, respectively, $\hat{r}$ is the longitudinal direction, and
the superscripts $a$ and $b$ indicate the two different perspectives. The $180^\circ$ ambiguity can be solved.
Moreover, a part of the transversal component of the magnetic field from one perspective is a part of
the longitudinal component from the other. The uncertainty of the transversal component therefore can be reduced
to some extent. Thus, multiple-perspective measurements will provide us unprecedented insight into the evolution
of the magnetic field from photosphere to corona.

But some issues exist and may affect the accuracy of the inversion of the magnetic field\citep{WangY_etal_2020a}. As indicated in
Figure~\ref{fg:bvec}, for example, the observational paths, heights of the emission source and the
optical depths from different perspectives are different though they aim to the same region. Thus,
the measured vector magnetic field of the same region could be theoretically different. The significance
of these effects depends on the separation angle of the two spacecraft. Obviously, the less the separation
angle is, the weaker are the effects. But if the separation angle is too small, the dual perspectives will
reduce to a single perspective considering the uncertainty in the measurements. Thus, the question is what the optimal
separation angle is between two spacecraft. In the current design, the separation angle is $30^\circ$ in
each group and $90^\circ$ between groups as illustrated in Figure~\ref{fg:sc_config}b. Intuitively, the $30^\circ$
separation is better than the $90^\circ$ separation: not only the inversion of the vector magnetic field
could be more accurate, but also the overlapped solar surface is wider.

It is worth to note that SolO by European Space Agency (ESA) was successfully launched on 10 February 2020.
It together with SDO at Earth will achieve the dual-perspective magnetic field observation for the first time. Since the
orbit of SolO is not in the ecliptic plane but has the inclination of about $25^\circ$ and perihelion of about $0.28$ AU,
the separation angle between SolO and SDO varies greatly. This causes that the accurate vector magnetic fields
can not be routinely obtained, which will be regret for the study of the global evolution or that of a particular region
not within the time window of the dual-perspective observations. But they will provide us opportunities to study the influence
of the separation angle on the inversion of the vector magnetic fields based on measured data. Besides, SolO gives the
perspective out of the ecliptic plane, which will be valuable addition to our Solar Ring mission. Combining the measurements
of the magnetic fields from three perspectives not lying on one plane, we theoretically can obtain vector magnetic fields
only based on the observations of the longitudinal components, which are more accurate than transversal components, as follows
\begin{eqnarray}
\left\{\begin{array}{l}
\vec{M}^a_L\vec{B}=B^a_L\\
\vec{M}^b_L\vec{B}=B^b_L\\
\vec{M}^c_L\vec{B}=B^c_L
\end{array}\right.
\end{eqnarray}
in which $\vec{M}_L$ is the longitudinal component of the transform matrix $\mathbb{M}$. The inversion of the vector
magnetic field should be even more accurate.

\paragraph{Provide 360-degree maps of the Sun and the inner heliosphere\\}

\begin{figure*}[t]
\centering
\includegraphics[width=0.85\hsize]{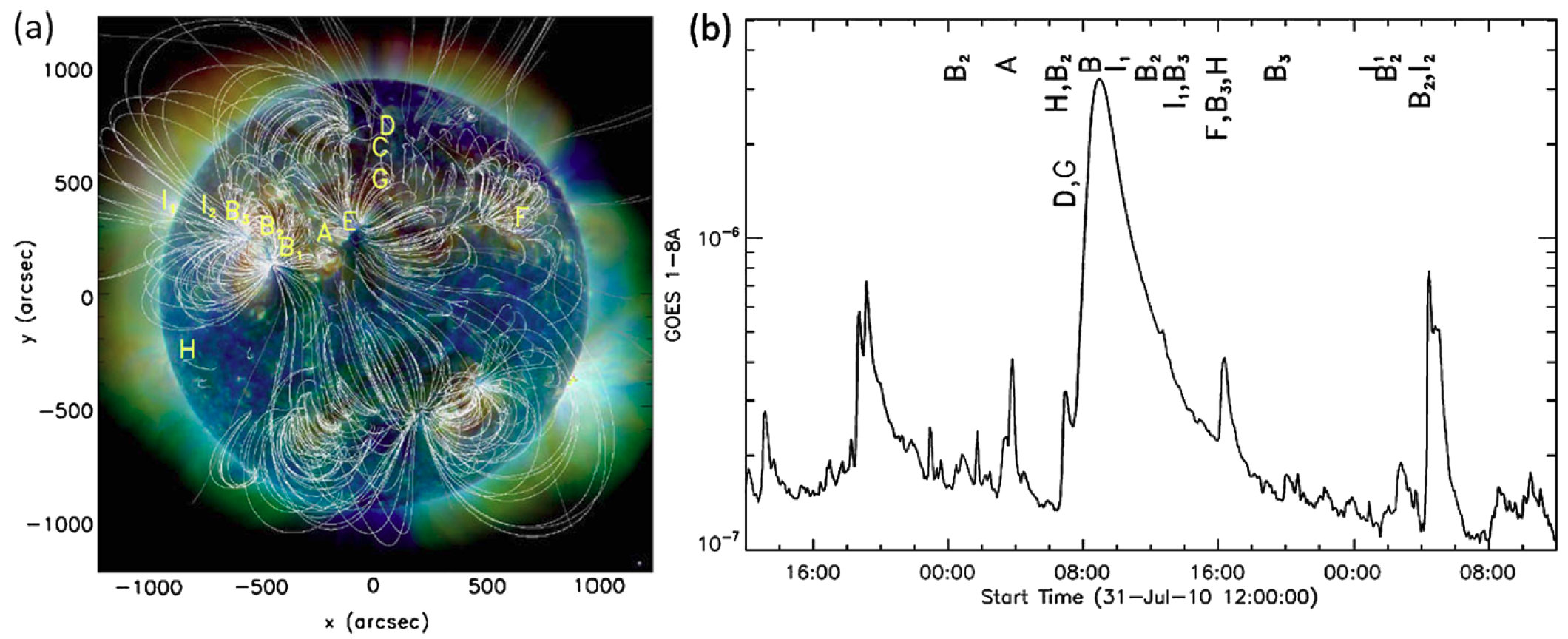}
\caption{{\it Panle (a):} Three-color composite EUV image combined from SDO/AIA 211\AA, 193\AA, and 171\AA channels
on 1 August 2010. Coronal magnetic field lines extrapolated using a potential field source surface (PFSS)
model are superimposed, showing the magnetic connections among different regions. Letters denote the locations
of the eruptive events during 1--2 August 2010. {\it Panel (b):} GOES 1--8\AA\ light curve with the same denoted
letters. Adapted from the paper\citep{Schrijver_Title_2011}.}
\label{fg:symp_events}
\end{figure*}

Multiple-perspective observations of global oscillations, which can be obtained when measuring the photospheric vector
magnetic fields, are especially important to helioseismology for studying the Sun's deeper interior
and higher latitude. Currently, all our knowledge on the Sun's interior structure and dynamics either relies on
theoretical modeling\cite{Christensen-Dalsgaard_etal_1996} or one-side observations of the Sun. The one-side
helioseismic observations, which routinely
started in mid-1990s\cite{Scherrer_etal_1995, Harvey_etal_1996}, allow us to infer the Sun's internal differential
rotation up to the bottom of the
convection zone\cite{Thompson_etal_1996, Howe_etal_2000}, and also allow us to start understanding the meridional
circulation inside the Sun\cite{ZhaoJ_etal_2013}. However,
what the differential rotation is like beneath the convection zone and in higher latitude above $60^\circ$ is largely
unknown. More importantly, despite the crucial role of meridional circulation in transporting magnetic flux inside
the Sun, our current understanding of the meridional-circulation structure is by far unsatisfactory. All these are
limited by our limited capability of only simultaneously observing limited areas of the Sun, with acceptable spatial
resolutions only within about $60^\circ$ from the Sun's apparent disk center. Solar Ring mission will allow us to
simultaneously observe the Sun from multiple perspectives with various angles including some greater than $120^\circ$,
providing crucial data for the deeper interior as well as higher latitude. A well-determined solar interior rotational
profile and meridional-circulation profile will allow us to better understand the Sun's dynamo and its generation of
magnetic cycles\cite{Miesch_Brown_2012}.

On the other hand, most notable solar activities are global behaviors. Sympathetic eruptions, for example, are often
observed\citep{Simnett_Hudson_1997, Moon_etal_2003, ZhouG_etal_2007}, which occur in different ARs but almost
simultaneously, suggesting connections and interactions between different ARs. Such connections
and interactions could be global and not just between neighboring regions. As a case studied in the
paper\citep{Schrijver_Title_2011}, the major events during 2011 August 1 -- 2, spreading over more than a
quarter of the solar surface in longitude, were connected via large-scale separators, separatrices and
quasi-separatrix layers (see Fig.\ref{fg:symp_events}). Even in single eruptions, inter-region connections
could be often identified\citep{ZhangY_etal_2007}. Statistical study also showed that one third of
all ARs present transequatorial loops\citep{Pevtsov_2000}. These facts require a $360^\circ$-view
of the Sun to completely and correctly understand such large-scale eruptive phenomena, including their causes
and effects.

Not only such widely-spread sympathetic activities, but also long-term (weeks to months) evolving solar
structures and features, e.g., filaments, sunspots, ARs and CHs, require the $360^\circ$ view.
For instance, the long-lived CH investigated in the paper\citep{Heinemann_etal_2016}
evolved for $10$ solar rotations from its growing phase to its maximum and decaying phases.
All these long-term evolving structures are controlled by the evolution of magnetic field and essentially
by the solar dynamo. So far the global photospheric magnetic field map, so called synoptic chart\citep{LiuY_etal_2012a},
is a kind of summary map of the sampled photospheric magnetic field over a solar rotation. Global
extrapolation of coronal magnetic field is based on such a synoptic chart. Without a realtime map, some details
especially those on the backside of the solar disk will be missed and the extrapolation will be inaccurate.

\begin{figure}[H]
\centering
\includegraphics[width=\hsize]{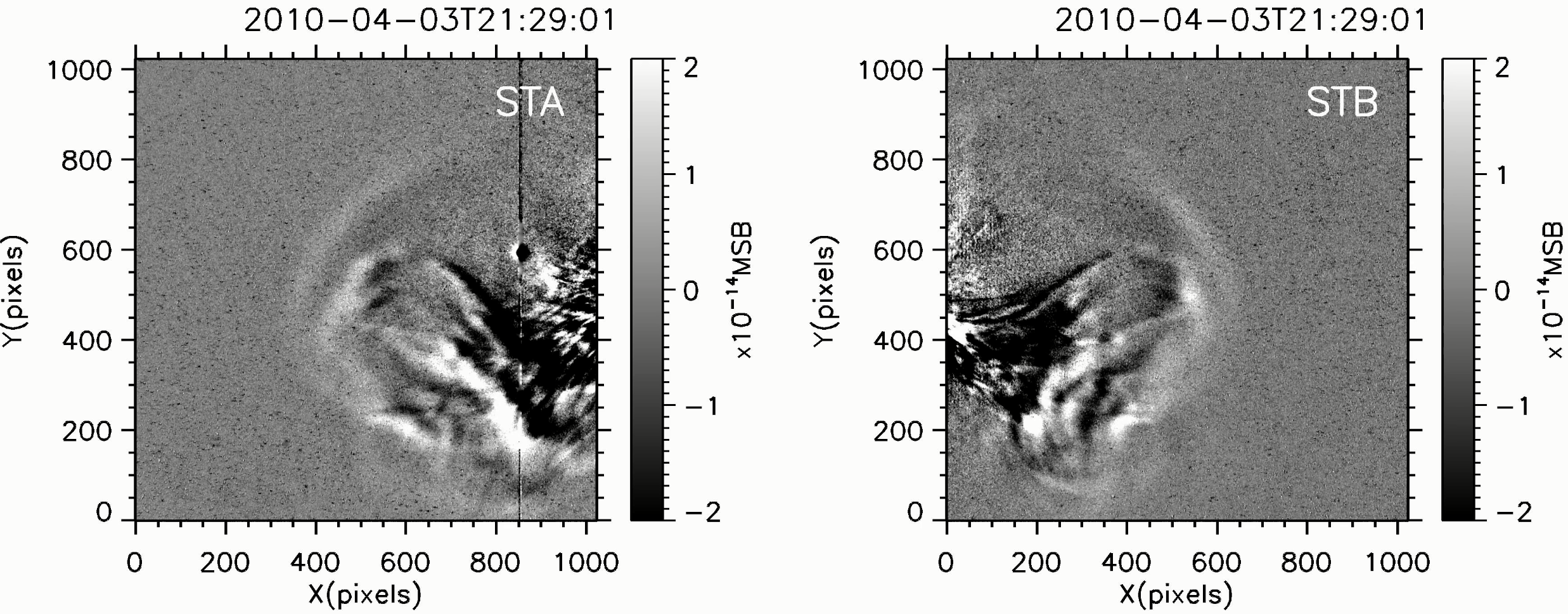}\\
\includegraphics[width=0.85\hsize]{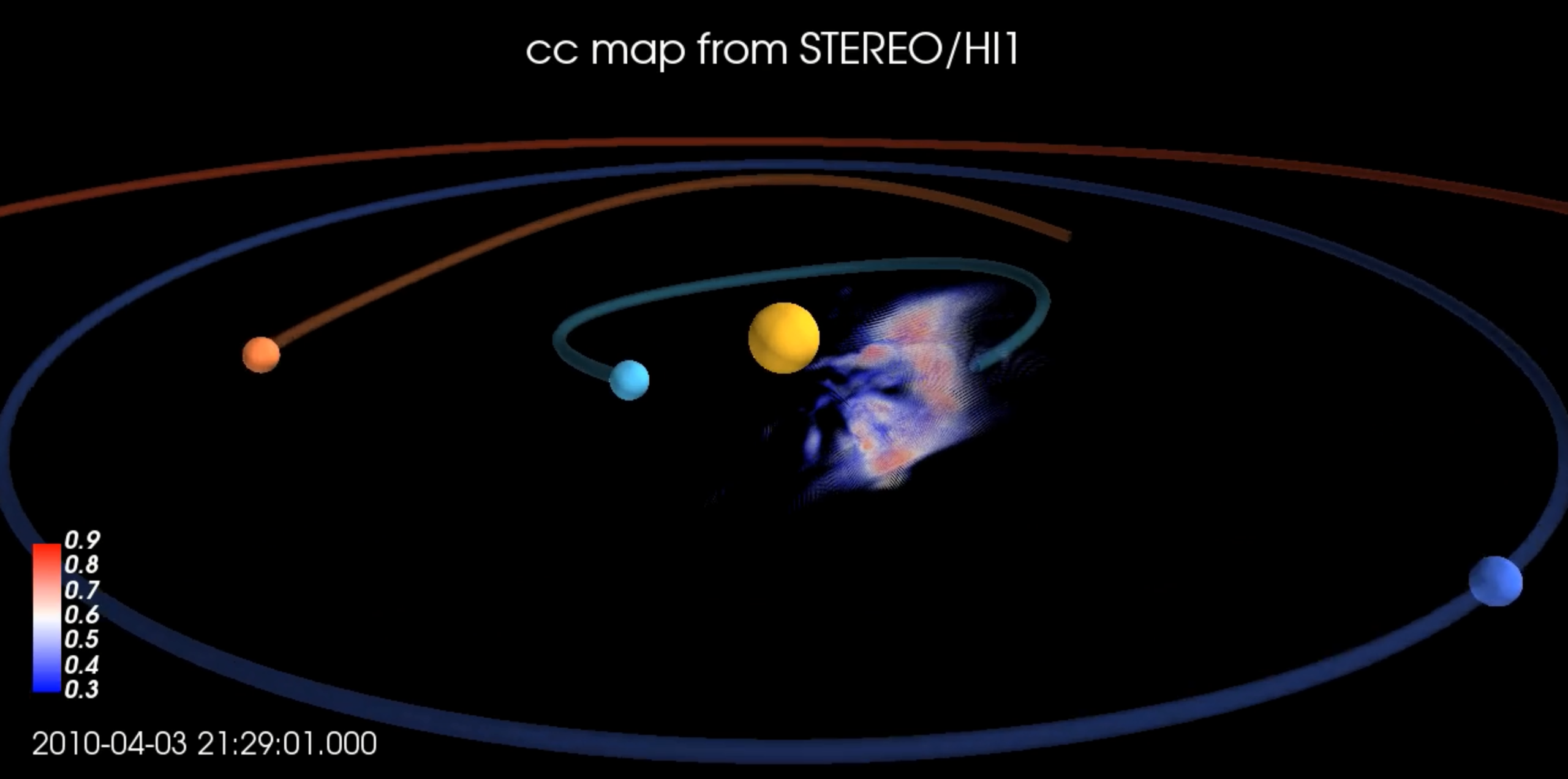}
\caption{{\it Upper Panels:} The running-difference images of the heliosphere taken by HI-1
cameras on board the STEREO A and B spacecraft at 21:29 UT on 3 April 2010. A CME was captured.
{\it Lower Panel:} The correlation coefficient (cc) map of the heliosphere at the same time, inferred from the
HI-1 images through CORAR method. The CME is reconstructed in the high cc region. The yellow, cyan, orange
and blue balls denote the Sun, Mercury, Venus and Earth. Adapted from the paper\citep{LiX_etal_2020}.}
\label{fg:hi1cme}
\end{figure}

From the $360^\circ$ maps of the Sun from multiple perspectives, we may also derive the 3D information of various
structures and features in the solar atmosphere, such as
the coronal loops\citep{Aschwanden_etal_2008}, solar jets\citep{LiuJ_etal_2014}, bright points\citep{Kwon_etal_2010},
etc. As illustrated in Figure~\ref{fg:sc_config}b, each group of the spacecraft may cover nearly $150^\circ$ in
longitude, and three spacecraft may cover $180^\circ$. Of course, the inversion of the 3D magnetic field and
coronal structures will become unreliable close to the edge of the common field of view (FOV). Thus, $150^\circ$
and $180^\circ$ are the upper limits in theory. But these gaps can be filled by other three spacecraft in
the Solar Ring mission. Then we can do what was never done before, e.g., to link different eruptive activities
scattered over the solar surface into one story, to trace the whole life of an
AR or a CH from its birth to death at multiple wavelengths, to trace how magnetic fluxes
transport from low latitude to high latitude and vise versa, and to forecast potential space weather events in
much more advance.

Between the Sun and planets including the Earth, there is a big gap --- interplanetary space or the heliosphere,
where the solar wind and various eruptive transients travel through all the time.
From the perspective of Earth, Earth-directing CMEs that most likely to cause severe space weather events might
be missed due to projection effect\citep{Robbrecht_etal_2009}. A statistical study suggested that about one third
of frontside CMEs were missed by the coronagraph LASCO on board SOHO\cite{Wang_etal_2011}. STEREO watching the corona
and inner heliosphere from two perspectives greatly reduce the projection effect, and make most Earth-directing
CMEs visible. Moreover, the 3D morphology and kinematics of various solar wind transients, including CMEs, blobs
and shocks, are able to be obtained from the dual-perspective imaging data\citep{Thernisien_etal_2006,
Sheeley_etal_2009, Lugaz_etal_2009, FengL_etal_2013}. The recent method based on Correlation-Aided Reconstruction (CORAR)
technique\citep{LiX_etal_2018, LiX_etal_2020} allows us to automatically recognize and locate inhomogeneous
structures in solar wind without any preset assumptions on the morphology. Figure~\ref{fg:hi1cme} shows
a CME (and three blob-like transients, not shown in the frame) propagating in the common
FOV of HI-1 cameras of STEREO on 2010 April 3 and the reconstruction by the CORAR technique, which is
quite consistent with the observations and other models\citep{LiX_etal_2020}. However, the common FOV of
STEREO is limited to the region between the Sun and Earth. Some large-scale structures often exceed or even travel
beyond the common FOV though they might also impact and affect the Earth's space environment\citep{Wang_etal_2016a}.
Thus, to have a panoramic 3D view of the inner heliosphere, more spacecraft, like the deployment of Solar Ring, is necessary.
A more detailed analysis of the optimal separation angle between the spacecraft for the the CORAR technique is
given in the paper\citep{Lyu_etal_2020}.

A panoramic view of the inner heliosphere is important to understand the dynamic evolution of solar wind
transients in interplanetary space and consequent space weather effects.
The outstanding questions include how the morphology and trajectory of solar wind transients change during the
propagation\citep{Wang_etal_2004b, Riley_Crooker_2004, Manchester_etal_2004, Wang_etal_2014, Kay_Opher_2015},
how the transients are accelerated or decelerated\citep{Gopalswamy_etal_2000, Vrsnak_etal_2008, Vrsnak_etal_2013,
Shen_etal_2014}, how the transients exchange magnetic flux with ambient solar wind\citep{Dasso_etal_2006, Ruffenach_etal_2015,
Wang_etal_2018}, how the transients interact with each other and cause the changes in velocity and direction\citep{Shen_etal_2012,
Lugaz_etal_2012, Temmer_etal_2014, Mishra_etal_2017}, etc. All these are closely-related to and partially determine whether,
when and how significantly severe space weather will occur.

\paragraph{Resolve solar wind structures at multiple scales and multiple longitudes\\}

Solar wind structures including steady structures, e.g., heliospheric current sheets and corotating
interaction regions (CIRs) forming between fast and slow solar wind, and aforementioned transients, e.g.,
blobs, CMEs, magnetic clouds and shocks, originate from the Sun and gradually propagate and expand into the heliosphere.
The macroscopic scale of them is very large. CMEs and magnetic clouds are all thought to be magnetic flux ropes
with their two legs still connected to the Sun even at the distance of $1$ AU\cite{Larson_etal_1997}; the angular width of them is
typically $60^\circ$\citep{Yashiro_etal_2004, Wang_etal_2011}, and the radius of the cross-section is about
$0.1$ AU on average at the heliocentric distance of 1 AU\citep{Wang_etal_2015}.

\begin{figure}[H]
\centering
\includegraphics[width=\hsize]{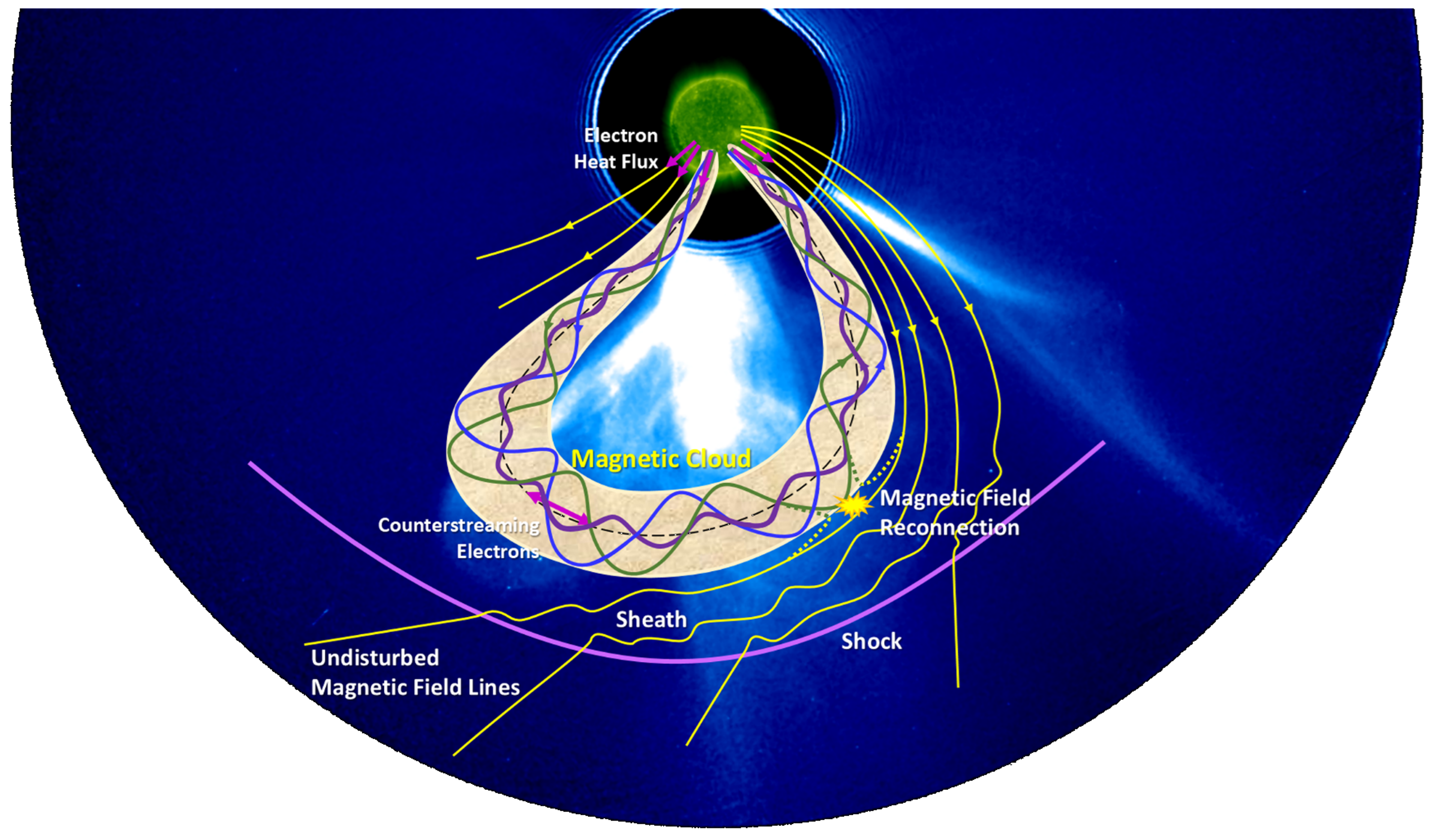}
\caption{A cartoon showing the large-scale magnetic flux rope, or called magnetic cloud,
in the heliosphere. The magnetic field lines are twisted in the magnetic cloud as indicated by the color-coded lines.
The reconnection site implies the erosion process. A shock exists if the magnetic cloud propagates fast. Adapted from the
paper\citep{Wang_etal_2018}.}
\label{fg:mfr}
\end{figure}

\begin{figure*}[t]
\centering
\includegraphics[width=\hsize]{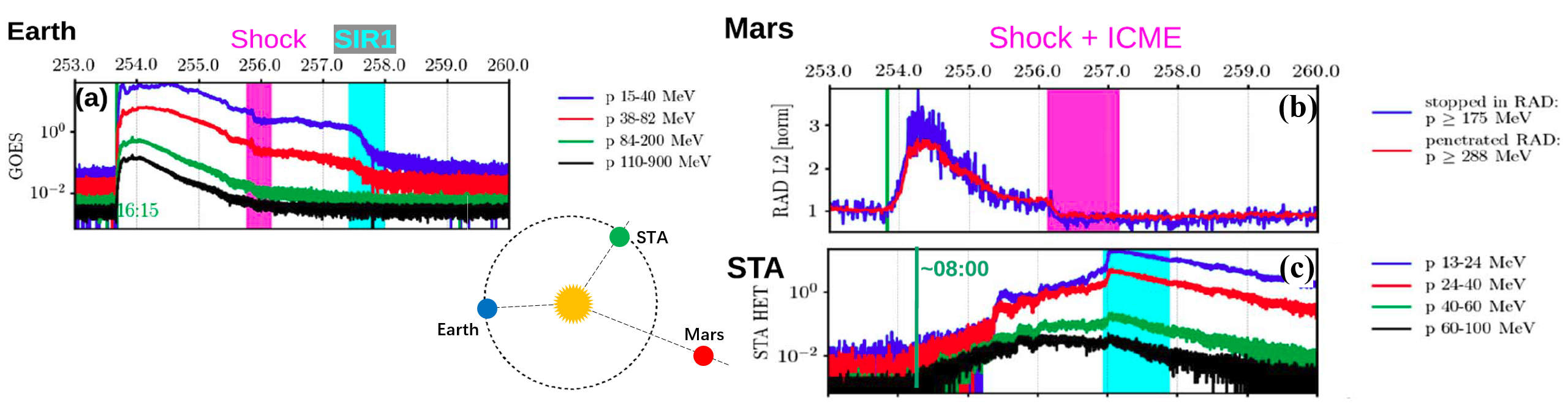}
\caption{The fluxes of energetic particles recorded by ({\it Panel a}) GOES at Earth, ({\it Panel b}) Radiation
Assessment Detector (RAD\citep{Hassler_etal_2012}) at Mars
and ({\it Panel c}) STEREO A during 10--16 September 2017. The pink region indicates the shock (and CME) arrival,
 and the cyan region the stream interaction region. The positions of the Earth, Mars and STEREO A on 10 September 2017
 are plotted too. A complex solar eruption caused the enhancement of the
fluxes in a wide range. Adapted from the paper\citep{GuoJ_etal_2018a}.}
\label{fg:sep}
\end{figure*}

Such a large-scale structure may manifest different properties and behaviors at different parts. Magnetic flux
rope, for example, is a fundamental plasma structure, and twist of magnetic field lines inside a interplanetary flux rope
is a key parameter to characterize its property and stableness and to understand its origin and initiation from the
Sun\citep{Wang_etal_2016, Wang_etal_2018} (Fig.\ref{fg:mfr}). However, it is not clear if the twist is uniformly distributed in
flux ropes. It is a debate whether or not the field lines are more twisted in the leg than in the apex of
a flux rope\citep{Demoulin_etal_2016, Wang_etal_2016, Owens_2016}. Moreover, magnetic clouds are thought
to be a coherent flux rope structure originating from the Sun, but it is difficult to identify the same
magnetic cloud structure from different spacecraft if the spacecraft separate too widely\citep{ZhaoA_etal_2017}.
Is it due to the loss of the coherency during the propagation in interplanetary
space\citep{Owens_etal_2017} or the strong modulation by ambient solar wind?

Interplanetary solar energetic particles (SEPs), with energies from a few keV up to relativistic GeV
are generally accelerated during solar eruptions by either the magnetic reconnection processes and/or CME-driven shocks.
SEPs generated by shocks are of particular interest since protons can reach energies larger than 10 MeV
posing serious radiation threats to human exploration activities in space and causing technological and
communication issues to satellites\citep{Desai_Giacalone_2016}. However, we have not yet obtained a complete understanding
of these large SEP events since their properties, normally observed from one viewpoint of Earth, are a complex mixture
of several important physical processes: acceleration, injection and transport. These processes are evolving in time
and location-dependent and are determined by the macro-structures of the shock and the heliosphere as well as the
micro-properties of the particle scattering and transport procedure. As a consequence, the distribution of
the flux of the energetic particles at different heliospheric longitude and radial distance may be quite
different\citep{Cane_etal_1988}, causing significantly different radiation environment at different planets.
A recent work using multiple spacecraft showed that a complex CME structure led to a solar energetic particle event
not only at Earth but also at Mars and STEREO A,
across a heliospheric longitude span of $230$ degrees\citep{GuoJ_etal_2018a}. However, SEP energy spectra and temporal
evolution are different at each of the three observers as shown in Figure~\ref{fg:sep}.
It is an interesting and important issue how energetic particles transport from
initial direction to such a wide longitude range.

Many macroscopic properties are linked with  microscopic processes, but in lack of details. For example, the
dramatic expansion of CMEs in interplanetary space may cause significant decrease of the internal temperature if no additional heat
source exists. However, the in-situ measurements at 1 AU show that the temperature is not low enough, suggesting
a heating process. With the aid of an observation constrained model, it was revealed for two CMEs that the
polytropic index of the CME plasma is well below the adiabatic index $3/5$, also suggesting the
heating\citep{Wang_etal_2009, Mishra_Wang_2018}. What is the mechanism of the heating? Is it due to the
wave-particle interaction or direct injection of thermal electrons?
Erosion of magnetic clouds has been proven a common phenomenon in interplanetary space\citep{Ruffenach_etal_2015}, suggesting a
strong exchange of the magnetic flux between the magnetic cloud and ambient solar wind. Its macroscopic manifestation is that a
magnetic cloud can be peeled off when it is propagating outward until merging into solar wind, while the microscopic manifestation
is the appearance of signatures of magnetic reconnection, i.e., the exhausting region, at the boundary of the magnetic
cloud\citep{Wang_etal_2010, Gosling_2012}. But without multiple spacecraft at different positions, it is hard to learn if the erosion
process is a local phenomenon or a global phenomenon, and if there are many exhausting regions spreading on the outmost surface of the
magnetic cloud. The same issue exists when studying the plasma motion inside magnetic clouds\citep{Wang_etal_2015, ZhaoA_etal_2017},
of which the cause is unclear.

Besides, the momentum transfer and energy conversion of solar wind transients with ambient
solar wind and other transients are still puzzling in many events. The collision/interaction between two CMEs in inner heliosphere
was never thought to be super-elastic before the 2008 November 2 event was analyzed based on imaging data\citep{Shen_etal_2012} and
numerical simulations\citep{ShenF_etal_2013}. However, due to the absence of the in-situ measurements of the event, the details of
the process and mechanism of how and why the total kinetic energy of the two CMEs was gained are still missing. Thus, in-situ measurements
at different positions as well as global pictures are required to better understand all of the above issues. The former is used
to obtain accurate plasma and magnetic field parameters, and the latter is to constrain global morphology of the large-scale structure,
which may provide additional information about the dynamic and thermodynamic processes of the structure, e.g., the acceleration/deceleration
process, pancaking or distortion process, erosion process, deflection process, interaction/collision process, etc.

There are events well observed by multiple spacecraft in both imaging data and in-situ measurements, but the number is small.
The configuration of the Solar Ring mission will make such kinds of observations routine. As mentioned
before, the angular width of CMEs are typically $60^\circ$\citep{Wang_etal_2011}, wider than the separation angle of the spacecraft
in each group of the Solar Ring mission. Shocks and resulted particle events could span even wider, and therefore the spacecraft belonging to
different groups with the separation angle of $90^\circ$ or $120^\circ$ could be ideal probes to measure them simultaneously. Other spacecraft
then take pictures from side-view to provide global information of the events. With this capability, the understanding of the solar wind structures
and the level of space weather forecasting will be greatly advanced.

\section{Scientific instruments and requirements}\label{sec:payload}

To achieve the above capabilities in accurately measuring photospheric vector magnetic fields, providing $360^\circ$
panoramic pictures of the Sun and the inner heliosphere, and resolving solar wind structures at multiple scales, and then
further to tackle the science objectives, we need remote sensing data, including spectral observations for the solar magnetic fields,
multi-band observations for the solar EUV emissions, white-light observations for the corona and inner heliosphere, and the
radio emissions, and also in-situ measurements of the solar wind magnetic field, solar wind plasma and energetic particles.
The relationship between the science objectives and these measurements is summarized in Table~\ref{tb:mea}.

\begin{table*}[tb]
\footnotesize
\caption{Science objectives and required measurements}\label{tb:mea}
\begin{tabular}{p{33px}|p{90px}|p{90px}|p{26px}|p{24px}|p{24px}|p{24px}|p{24px}|p{24px}|p{24px}}
\hline
Science objectives & Scientific questions & Strategy & Solar magnetic field \& global Doppler velocity & Solar EUV images & White-light images & Radio emissions & Solar wind magnetic field & Solar wind plasma & Energetic particles \\
\hline
Origin of solar cycle & How does the global magnetic flux emerge, transport and dissipate? & Trace the global magnetic fluxes at multiple scales. & $\surd$ & & & & & & \\
& What is the solar internal structure? & Analyze the global oscillation modes. & $\surd$ & & & & & & \\
&&&&&&&&& \\
Origin of solar eruptions & How is the energy accumulated and released, and how is an eruption triggered? & Trace the evolution of source region and combine measured magnetic field, radio emissions and energetic particles to estimate some key parameters, e.g., the magnetic energy and helicity, and key processes. & $\surd$ & $\surd$ & & $\surd$ & & & $\surd$ \\
& How are the coronal structures reconstructed, and what kind of structures are formed and ejected into heliosphere? & Extrapolate coronal field and compare with observed coronal plasma structures in EUV, compare erupted signatures in EUV and white-light images. & $\surd$ & $\surd$ & $\surd$ & & & & \\
&&&&&&&&& \\
Origin of solar wind structures & Where does an solar wind structure come from? What's its topology and magnetic connection with the Sun? How does a solar wind structure evolve in the heliosphere in terms of its propagation direction, velocity, topology, etc? & Use white-light images from multiple perspectives to recognize and reconstruct solar wind structures in interplanetary space and trace their evolution. Associate the imaging data of solar wind structures to in-situ data at 1 AU to confirm their properties, and trace back to the Sun to obtain the properties of their sources.  & $\surd$ & $\surd$ & $\surd$ & $\surd$ & $\surd$ & $\surd$ & $\surd$ \\
&&&&&&&&& \\
Origin of severe space weather events & What are the primary factors causing major geomagnetic storms and/or solar energetic particle events? & Investigate in-situ data, including magnetic field, solar wind plasma and energetic particles, at different longitudes to assess the effects of various factors on the space weather. & & & $\surd$ & & $\surd$ & $\surd$ & $\surd$ \\
& What are the properties of the source regions of the drivers of severe space weather? How can we make an accurate forecast of the space weather effects of solar eruptions? & Use imaging data of the heliosphere and the Sun to identify the source regions of the space-weather-effecting solar wind structures and to study the relationship between the solar eruptions and space weather events. & $\surd$ & $\surd$ & $\surd$ & $\surd$ & $\surd$ & $\surd$ & $\surd$ \\
\hline
\end{tabular}\\
\end{table*}

These measurements are suggested to be accomplished by the following science payloads onboard each of
the spacecraft of the Solar Ring mission (refer to Table~\ref{tb:tech1}). First of all, the Spectral Imager for Magnetic field and
helioSeismology (SIMS) provides the measurements of the vector magnetic field and Doppler shift information
on the photosphere. It is one of the most important payloads regarding the four science objectives: the origin of
the solar cycle, the origin of the solar eruptions, the origin of the solar wind structures and the origin of the
severe space weather events. SDO/HMI can provide the $4096\times4096$ vector magnetogram of the full photosphere every $45$ s in the
longitudinal component and every $135$ s in the transversal component. Such cadences and spatial resolution are able to reveal
the magnetic evolution and therefore the accumulation process of magnetic energy before an eruption, and are sufficient for the
study of the solar cycle. Considering the orbit of the designed Solar Ring spacecraft, which will be discussed in the next section,
being much farther than that of SDO and limiting the data transmission rate, the cadence of SIMS
is set to close to that of HMI when the spacecraft close to the Earth and $1$ hr or longer at far side.

The Multi-band Imager for EUV emissions (MIE) provides the condition and evolution of the plasma structures in the
solar atmosphere, where are necessary to see the effect of the evolution of photospheric magnetic field. The spatial
resolution is the same as SIMS. With images from
multiple perspectives, the 3D topology of plasma structures can be revealed, of which the process is similar to those done
by using STEREO and/or SDO data\citep{Aschwanden_etal_2008}.
These data provided by MIE are key to understand the eruptive phenomena in the solar corona
and also to locate the solar source of the solar wind structures traveling in the outer corona and inner heliosphere. Three
wavelength bands are suggested: (1) $304\AA$, a relative cool line for the chromosphere and transition region, particularly
suitable for filaments/prominences, (2) $171\AA$, a relatively warm line for the corona, good for coronal loops, post-flare
arcades, etc., and (3) $131\AA$, a relatively hot line for flaring regions with a warm component less than
$1$ MK, best for hot channels and other heated structures.

The Wide-Angle Coronagraph (WAC) provides the situation of the outer corona through inner heliosphere, bridging the Sun
and inner planets, including Mercury, Venus, Earth and probably Mars. The white-light images taken by WAC are necessary
to identify the consequence of the solar eruptions in interplanetary space and
the source of space weather events. The data from multiple spacecraft can be further used to retrieve the 3D information of
solar wind structures. The brightness of solar wind decreases quickly with increasing distance away from the Sun,
causing strong contrast between near-Sun side and far-side. The signal-to-noise ratio decreases too with increasing distance,
and particularly beyond elongation angle of $20^\circ$, it becomes too
low to perform a reliable reconstruction\citep{LiX_etal_2020}. To reach a compromise among
the wide FOV, the acceptable contrast and the sufficient signal-to-noise, WAC is suggested to have an outer FOV of about
$\pm12^\circ$ in elongation angle and an inner FOV of about $\pm2^\circ$, covering the region from about $7.5$ to
$45 R_S$ in the plane-of-the-sky from $1$ AU (or about $5.6$ to
$34 R_S$ from $0.75$ AU). This region is best for the study of solar wind transients in 3D as most of them have
been well developed and entered the cruise phase. Due to the large scale of the structures, the cadence of the images could
be tens of minutes depending on the signal-to-noise and the data transmission rate.

The radio investigator (WAVES) is deployed to monitor the high-energy phenomena occurring on the Sun and in interplanetary
space. Solar flares will cause Type III and Type IV radio bursts, and fast-forward shocks driven by CMEs will cause Type IIs.
These bursts leave distinguished drift patterns in the radio dynamic spectrum from GHz to below MHz, providing additional diagnose of
solar eruptions. The WAVES is preliminarily designed to receive the radio signal from $30$ MHz to $5$ kHz, slightly wider than
Wind/WAVES and STEREO/WAVES, covering the heliocentric distance from about $1.5 R_S$ to $1$ AU in terms of the electron plasma frequency.
When multiple spacecraft receive the same burst, its radio source region could be located by using a triangulation
method\citep{Reiner_Stone_1986, Krupar_etal_2014, Magdalenic_etal_2014, ZhangP_etal_2019}. The method in the paper\citep{ZhangP_etal_2019}
showed that the 1-minute temporal resolution of the radio intensity spectrum can be used to locate the source about $25 R_S$ away
from the Sun if the spacecraft are separated by more than $60^\circ$. The wider the spacecraft are separated, the lower temporal
resolution is sufficient. Thus, we suggest the temporal resolution of final radio dynamic spectrum should be better than $30$ s
to locate the source region closer to the Sun.
Besides, more information of the radio emission, e.g., the polarization properties, need the instantaneous goniopolarimetric (GP)
capability (also know as direction-finding capability) of the radio receivers\citep{Cecconi_etal_2008}. It requires rapid switch
between the channel/antenna configurations about every $0.2$ s, which should also be equipped. This capability will also increase
the accuracy in locating the source region of a radio burst, benefitting the space weather forecasting.

The rest of the payloads are for in-situ measurements to provide a more complete picture of the conditions of the inner heliosphere.
As one of the fundamental parameters characterizing the solar wind conditions, the interplanetary magnetic field is typically
$10$ nT on average, and sometimes can reach up to hundreds of nT. For solar
wind transients which are large-scale structures, the sampling rate of magnetic field does not need to be too high.
But for microscopic phenomena and process, e.g., the shock front, reconnection exhausting region and turbulence, a high sampling
rate of magnetic field is required. The previous studies have shown that the wave energy carried by the solar wind cascades from
large scale to small scale in inertial range of the wavelength, and reaches the dissipation range, which is typically beyond
$1$ Hz\citep{Leamon_etal_1998}. Thus, the Flux-Gate Magnetometer (FGM) is suggested to be deployed to sample the magnetic field
with the rate of $0.1$ or $128$ Hz (depending on the data transmission rate) and the resolution of $0.01$ nT.

The Solar wind Plasma Analyzer (SPA) measures the in-situ plasma in the energy range from $0.1$ to $25$ keV for ions and $0.05$ to
$10$ keV for electrons. This energy range covers the main flux of the solar wind plasma, providing other basic parameters, e.g.,
the bulk velocity, density and temperature, of the solar wind conditions. Solar wind outflow has two components. One
is the beam almost along the Sun-observer line, and the other the halo mostly coming from the Sun-ward half sky due to the
scattering and diffusion. Thus, the SPA is designed to receive particles with the FOV of $180^\circ$ (azimuthal angle)
$\times\pm45^\circ$ (polar angle)\cite{HuR_etal_2019}. Further, to diagnose the source of solar wind, e.g., CHs or ARs, steady flow or transients,
the capability of measuring the mass and charge state is required to distinguish Helium, Carbon, Oxygen through Iron ions. The
temporal resolution ranges from seconds to minutes.

The High-energy Particle Detector (HiPD) is dedicated to the study of the particle acceleration and transportation and the solar
energetic particle events, related to three of the four science objectives as indicated in Table~\ref{tb:mea}.
The HiPD is designed similar to the High  Energy Telescope (HET) on board the SolO mission, consisting of four
$300$ um thick silicon solid state detectors (SSDs) and one high-density scintillation crystal. It measures electrons,
protons, and heavy ions. Electrons are covered across the energy range from $500$ keV up to about $20$ MeV; protons are
measured between $10$ and $100$ MeV and heavy ions from about $20$ to $200$ MeV/nuc.
HiPD also needs to separate $^3$He and $^4$He isotopes which is
important in differentiating the acceleration process of flare or shock related SEPs. During solar quiet times, the
study of these particles as present in galactic cosmic rays (GCRs) can also help to understand the transport of GCRs
inside the heliosphere. The HiPD unit will be located on each of the six Solar Ring spacecraft with the central
FOV pointing sunward/anti-sunward direction of the Parker Spiral and a view cone of about $55^\circ$.
The sunward direction will allow the detection of the beam SEPs which are the earliest arriving ones during an
event especially when the observer is well connected to the acceleration.
The anti-sunward direction will allow the detection of back scattered particles
which are excellent tracers of the magnetic topology and large scale connectivity of the interplanetary magnetic field\citep{Malandraki_etal_2005}.

In total, the mass of the suggested payloads of each spacecraft is $110$ kg, the power requirement is $180$ W, and the data rate at peak time is
$52.06$ Mbps. The data rate might be too high to achieve based on the current technology. Thus, how to reduce and compress the data
or to develop a new technique to communicate with the spacecraft is a big challenge, and will be studied in the future.

\begin{table*}[tb]
\footnotesize
\caption{Main tasks and preliminary technical specifications of payloads (to be continued)}\label{tb:tech1}
\begin{tabular}{p{125px}|p{145px}|l}
\hline
Payloads & Main tasks & Preliminary technical specifications \\
\hline
\multirow{8}{125px}[38px]{Spectral Imager for Magnetic field and helioSeismology (SIMS)} & \multirow{8}{145px}[28px]{Measure photospheric vector magnetic field to learn the global transportation of magnetic flux; measure global Doppler velocity to learn the global oscillations.} & Mass: $\leq30$ kg \\
& & Power consumption: $\leq40$ W \\
& & Data rate: $\leq30$ Mbps (@peak time) \\
& & Field of view: $32'\times32'$ (@1 AU) \\
& & Effective pixels: no less than $4096\times4096$ \\
& & Spectral resolution: better than $0.04\AA$ \\
& & Temporal resolution of longitudinal component: $1$ min $1$ hr \\
& & Temporal resolution of transversal component: $2$ min $1$ hr \\
&&\\
\multirow{7}{125px}[32px]{Multi-band Imager for EUV emissions (MIE)} & \multirow{7}{125px}[5px]{Obtain the global EUV images of solar disk at three wavelength bands, corresponding to relatively cool, warm and hot temperatures, respectively, to learn the morphology, topology, connectivity and emission measure of various plasma structures.} & Mass: $\leq30$ kg  \\
& & Power consumption: $\leq60$ W \\
& & Data rate: $\leq21$ Mbps (@peak time) \\
& & Field of view: $42'\times42'$ (@1 AU) \\
& & Effective pixels: no less than $4096\times4096$ \\
& & Wavelength bands: $304\AA$, $171\AA$, $131\AA$ \\
& & Temporal resolution: $10$ s, $1$ min, $1$ hr \\
&&\\
Wide-Angle Coronagraph (WAC) & \multirow{7}{125px}[11px]{Obtain white-light images of the solar wind structures traveling through the outer corona and inner heliosphere to learn their kinematic properties; get total brightness and the variations to learn the density distribution in 3D.} & Mass: $\leq25$ kg  \\
& & Power consumption: $\leq40$ W \\
& & Data rate: $<1$ Mbps (@peak time) \\
& & Field of view:  $\pm12^\circ$ \\
& & Occulting disk: $\pm2^\circ$ \\
& & Effective pixels: no less than $4096\times4096$ \\
& & Temporal resolution: $1$ min, $1$ hr \\
&&\\
Radio investigator (WAVES) & \multirow{6}{125px}[10px]{Measure the electric field intensity induced by the radio emissions from the Sun to recognize the radio bursts and get the location of the driving source and its kinematic properties.} & Mass: $\leq15$ kg\\
& & Power consumption: $\leq16$ W \\
& & Data rate: $0.5$ kbps \\
& & Frequency range: $5$ kHz -- $30$ MHz \\
& & Frequency channels: no less than 160 \\
& & Temporal resolution: better than $30$ s \\
& & GP mode: $0.2$ s for each channel/antenna configuration \\
\hline
\end{tabular}\\
\end{table*}

\begin{table*}[tb]
\footnotesize
\caption{Main tasks and preliminary technical specifications of payloads (continued)}\label{tb:tech2}
\begin{tabular}{p{125px}|p{145px}|l}
\hline
Payloads & Main tasks & Preliminary technical specifications \\
\hline
Flux-Gate Magnetometer (FGM) & \multirow{9}{125px}[34px]{Measure the in-situ magnetic field at 1 AU to learn the variations during solar wind structures and the distribution in longitude.} & Mass: $\leq2$ kg \\
& & Power consumption: $\leq3$ W \\
& & Data rate: $\leq8$ kbps \\
& & Maximum measuring range: $\pm65000$ nT \\
& & Dynamic measurement range: $2000$ nT \\
& & Resolution: better than $0.01$ nT \\
& & Noise level: better than $0.01$ nT/$\sqrt{Hz}$ \\
& & Zero drift: better than $0.01$ nT/$^\circ C$ \\
& & Sampling rate: $0.1$ Hz, $128$ Hz \\
&&\\
Solar wind Plasma Analyzer (SPA) & \multirow{16}{125px}[67px]{Measure the in-situ solar wind plasma at 1 AU; obtain the velocity, density, temperature and composition of the solar wind to learn the variations during solar wind structures and the distribution in longitude.} & Mass: $\leq7$ kg \\
& & Power consumption: $\leq 20$ W \\
& & Data rate: $\leq50$ kbps \\
& & Field of view: $180^\circ$ (azimuthal angle) $\times$ $\pm45^\circ$ (polar angle) \\
& & Angular resolution: better than $12^\circ$ (azimuthal) $\times$ $15^\circ$ (polar) \\
& & Temporal resolution: $4-64$ s (adjustable) \\
& & Ions \\
& & $\bullet$ Energy range: $0.1-25$ keV \\
& & $\bullet$ Energy resolution: better than 12\% \\
& & $\bullet$ Energy channels: no less than 64 \\
& & $\bullet$ Mass range: $0-60$ amu \\
& & $\bullet$ Mass resolution: better than 18\% \\
& & Electrons \\
& & $\bullet$ Energy range: $0.05-10$ keV \\
& & $\bullet$ Energy resolution: better than 12\% \\
& & $\bullet$ Energy channels: no less than 64 \\
&&\\
High-energy Particle Detector (HiPD) & \multirow{8}{125px}[22px]{Measure energetic particles in multiple energies to obtain the intensity and spectrum of a solar energetic particle event and to learn its driver and the distribution in longitude.} & Mass: $\leq1$ kg \\
& & Power consumption: $\leq 1$ W \\
& & Data rate: $\leq1$ kbps \\
& & Field of view: $55^\circ$ cone \\
& & Mass range: $0-60$ amu, electrons \\
& & Energy range of \\
& & $\bullet$ Electrons: $0.5-20$ MeV \\
& & $\bullet$ Protons: $10-100$ MeV \\
& & $\bullet$ Heavy ions: $20-200$ MeV/nuc \\
\hline
Total & \multicolumn{2}{l}{Mass: $\leq110$ kg, Power: $\leq180$ W, Data rate: $\leq52.06$ Mbps}\\
\hline
\end{tabular}\\
\end{table*}

\section{Mission profile and design}\label{sec:design}

There are several factors defining the basic outline of the Solar Ring mission. (1) The number and separation angles of the spacecraft.
As mentioned before, the number is six and the separation angles of them are about $30^\circ$ in each group and $120^\circ$ between groups as
shown in Figure~\ref{fg:sc_config}. The orbit is within $1$ AU, and could be a circular orbit or elliptical orbit.
A circular orbit may provide a more consistent FOV among the spacecraft than an elliptical orbit, and be
more convenient for and more accurate in the analysis of the data from multiple perspectives. However,
to deliver spacecraft into a circular orbit consumes more fuel than into an elliptical orbit.
(2) The time to deploy all the spacecraft. The science objectives require all the spacecraft working together.
For some objectives, at least two spacecraft are required. Considering the lifetime of each spacecraft,
it is better deploy all the spacecraft within a short time. The scheme of one rocket two spacecraft or even four
spacecraft can deploy the mission fast, but needs larger rocket thrust that reduces the capability of carrying payload.
(3) The mass of the spacecraft. The dry mass of a spacecraft without payloads is at least $350$ kg. Plus the suggested
payloads, the total mass is about $460$ kg. This also defines the type of rocket that is suitable for the mission.
(4) The cost of the launch. A cheaper cost makes the mission more feasible.

\begin{figure}[H]
\centering
\includegraphics[width=\hsize]{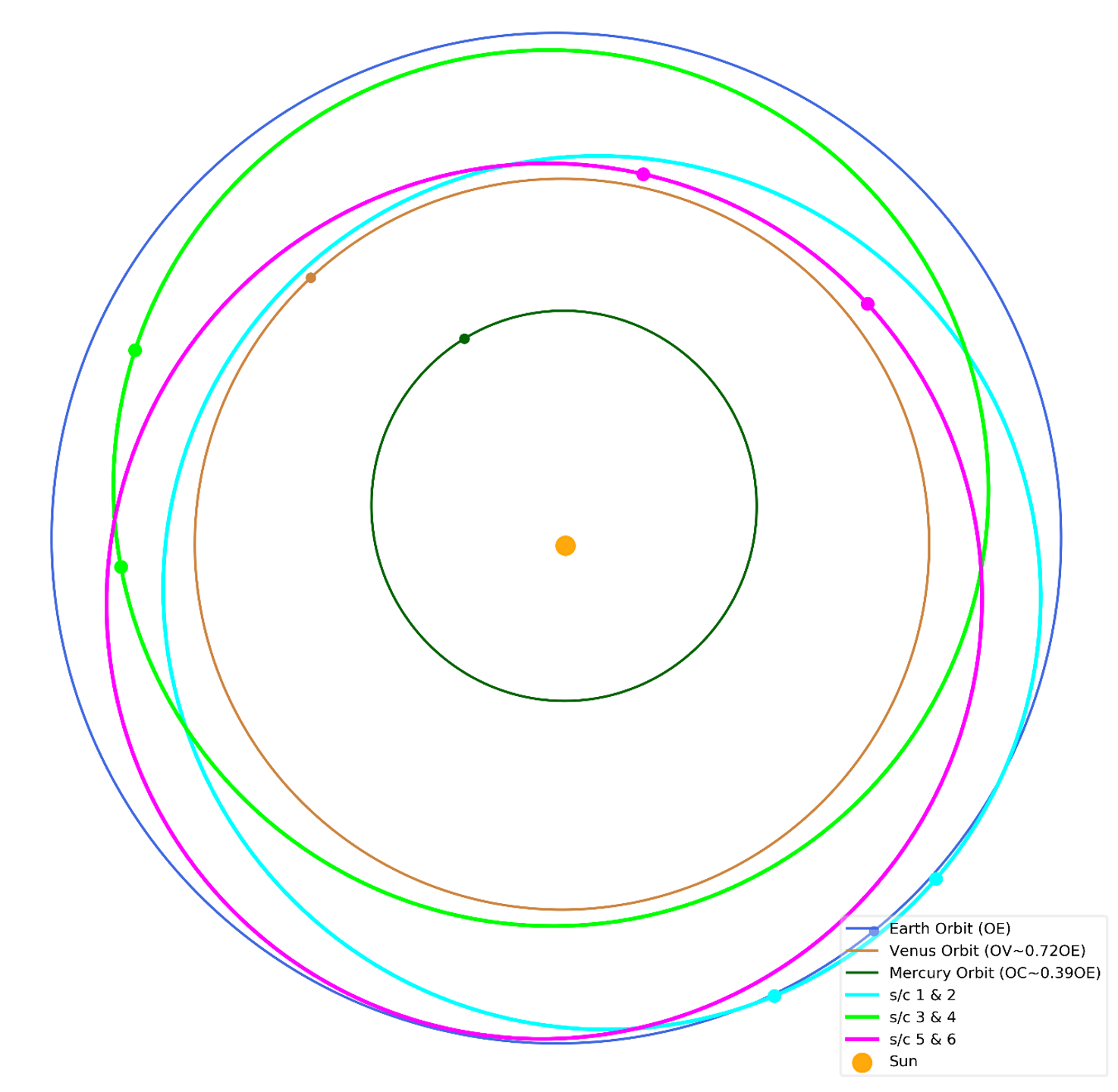}
\caption{The elliptical orbits (cyan, light green and pink) of the three groups of the spacecraft with the perihelion of $0.75$ AU and the
aphelion of $1$ AU. In this scheme, the separation angle among the three groups oscillates around $120^\circ$ and the angle between
the spacecraft in each group oscillates around $30^\circ$.}
\label{fg:elliptical_orbit}
\end{figure}

Combining the above considerations, the circular orbit is not the best option.
Here we propose a low cost scheme (a more detailed and complete analysis
of the mission profile and design can be found in the companion paper\citep{WangY_etal_2020}).
The elliptical orbit with the perihelion between $0.7$ and $0.85$ AU and the aphelion at $1$ AU (Fig.\ref{fg:elliptical_orbit})
is preliminarily adopted.In this scheme, one rocket two spacecraft technology could be applied to shorten the deployment
time and save the launch cost. But a few fuel is needed for the second spacecraft to adjust the orbital phase to
accomplish the $30^\circ$ separation from the first spacecraft. The time for the orbital phase adjustment
is about one year, but really depending on the perihelion and the carrying capacity of the rocket.
The preferred rocket could be Long March 3A (LM-3A), which has the carrying capacity of about $1400$ kg and can carry two
spacecraft each time. To make an even faster deployment, Long March 3B (LM-3B), which has much larger carry capacity,
could be considered, but the cost is about two times of that of using LM-3A.

If choosing the elliptical orbit with the perihelion of $0.85$ AU, the spacecraft separate from the Earth at a speed of
about $47^\circ$ per year. After $2.54$ years, the second group of spacecraft could be launched to form $120^\circ$ separation
between the two groups. Similarly, the third group will be launched about $5.08$ years later since the launch of the first group. The final
configuration of the Solar Ring will form in about $6.5$ years. A smaller perihelion could be chosen to shorten the deployment time.
If the perihelion, for example, is $0.7$ AU, the total time to finish the deployment is about $3.65$ years, but the carrying
capacity of LM-3A may not be sufficient. Besides, it should be noted that the three groups of
the spacecraft are not on the same elliptical orbit in this launch scheme (Fig.\ref{fg:elliptical_orbit}),
and therefore the separation angles
among them will vary around the designed values with time. This is acceptable as our science objectives do not
require the fixed separation angles and heliocentric distances of the spacecraft.

\begin{figure}[H]
\centering
\includegraphics[width=\hsize]{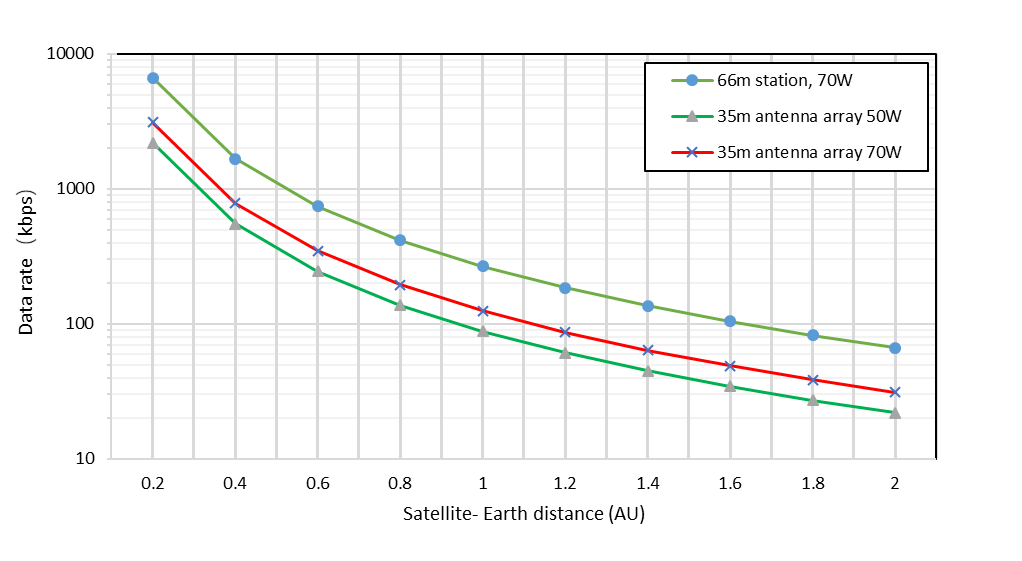}
\caption{The data transmission rate as a function of the distance between the spacecraft and Earth. Different lines
show the rate for the telescope with different size and different power.}
\label{fg:transmission}
\end{figure}

In this design, the distance between the spacecraft and Earth varies within $2$ AU. A larger antenna and power
is required to receive sufficient data. Assuming each spacecraft is equipped a communication antenna with the
aperture of $1.2$ m and supplied with the power of $70$ W, we have the data transmission rate from about $5000$ kbps
at 0.25 AU away from the Earth to less than $70$ kbps at $2$ AU away by using the $66$-m telescope at Jiamusi station
(see Fig.\ref{fg:transmission}). If we have $8$ hours for data transmission
every day, we can receive less than $20$ GB data per day at $0.25$ AU away or $200$ MB per day at $2$ AU away
for one spacecraft. It is much lower than the desired data rate based on the current payload requirement, becoming
the the strongest restriction to this mission. To solve or relieve this problem, either we reduce the data rate by enhancing
the capability of the onboard data processing, compression and storage and decreasing the sampling frequency, or
we develop more efficient techniques for the deep space communication, e.g., laser communication\cite{WuW_etal_2018}.

\section{Summary and conclusions}\label{sec:summary}

This ambitious concept of Solar Ring mission aims to achieve unprecedented capabilities to advance our understanding of
the Sun and the inner heliosphere from four aspects: the origin of solar cycle, the origin of solar eruptions,
the origin of solar wind structures and the origin of severe space weather events. The cost of the whole mission
is huge, but the design of the three groups of spacecraft makes the international collaboration being an option,
which may reduce the financial load of any single country. There are lots of challenges in the technique of carrier,
control, communication, payloads, etc, that need to be justified in the next study. We are looking forward to the
mission concept coming true.\\[5px]

\noindent{\it \footnotesize We thank Dr. J. Zhao from Stanford University for reading the manuscript and providing suggestions.
This work was supported by the Strategic Priority Program of CAS (Grant Nos. XDA15017300 and XDB41000000)
and the National Natural Science Foundation of China (NSFC, Grant No.41842037).
Y.W., C.S., J.G., Q.Z., K.L., X.L., R.L. and S.W. are also supported by the CAS (Grant No. QYZDB-SSW-DQC015) and the NSFC
(Grant Nos. 41774178, 41761134088, 41750110481 and 11925302), H.J. by the NSFC (Grant No. 11790302), and L.X. by the NSFC (Grant No. 41627806).
}

\noindent{\footnotesize {\bf Conflict of Interest} The authors declare that they have no conflict of interest.}

\bibliographystyle{unsrt}
\bibliography{../../ahareference}

\begin{thebibliography}{100}

\bibitem{Hudson_etal_2006}
H.~S. Hudson, J.~L. Bougeret, and J.~Burkepile.
\newblock Coronal mass ejections: Overview of observations.
\newblock {\em Space Sci. Rev.}, 123:13--30, 2006.

\bibitem{Yashiro_etal_2004}
S.~Yashiro, N.~Gopalswamy, G.~Michalek, O.~C. {St. Cyr}, S.~P. Plunkett, N.~B.
  Rich, and R.~A. Howard.
\newblock A catalog of white light coronal mass ejections observed by the soho
  spacecraft.
\newblock {\em J. Geophys. Res.: Space Phys.}, 109:A07105, 2004.

\bibitem{Dikpati_Charbonneau_1999}
Mausumi Dikpati and Paul Charbonneau.
\newblock A {Babcock-Leighton} flux transport dynamo with solar-like
  differential rotation.
\newblock {\em Astrophys. J.}, 518:508--520, 1999.

\bibitem{Reid_1999}
George~C. Reid.
\newblock Solar variability and its implications for the human environment.
\newblock {\em J. Atmos. Solar-Terres. Phys.}, 61:3--14, 1999.

\bibitem{Lean_Rind_1999}
Judith Lean and David Rind.
\newblock Evaluating sun–climate relationships since the {Little Ice Age}.
\newblock {\em J. Atmos. Solar-Terres. Phys.}, 61:25--36, 1999.

\bibitem{Nandy_etal_2011}
Dibyendu Nandy, Andr{\'e}s Mu{\~n}oz-Jaramillo, and Petrus C.~H. Martens.
\newblock The unusual minimum of sunspot cycle 23 caused by meridional plasma
  flow variations.
\newblock {\em Nature}, 471:80--82, 2011.

\bibitem{Schrijver_etal_2011a}
Carolus~J. Schrijver, W.~C. Livingston, T.~N. Woods, and R.~A. Mewaldt.
\newblock The minimal solar activity in 2008-2009 and its implications for
  long-term climate modeling.
\newblock {\em Geophys. Res. Lett.}, 38:L06701, 2011.

\bibitem{McComas_etal_2013}
D.~J. McComas, N.~Angold, H.~A. Elliott, G.~Livadiotis, N.~A. Schwadron, R.~M.
  Skoug, and C.~W. Smith.
\newblock Weakest solar wind of the space age and the current "mini" solar
  maximum.
\newblock {\em Astrophys. J.}, 779:2(10pp), 2013.

\bibitem{Feulner_Rahmstorf_2010}
Georg Feulner and Stefan Rahmstorf.
\newblock On the effect of a new grand minimum of solar activity on the future
  climate on earth.
\newblock {\em Geophys. Res. Lett.}, 37:L05707, 2010.

\bibitem{Domingo_etal_1995}
V.~Domingo, B.~Fleck, and A.~I. Poland.
\newblock {SOHO}: The solar and heliospheric observatory.
\newblock {\em Space Sci. Rev.}, 72:81--84, 1995.

\bibitem{Handy_etal_1999}
B.~N. Handy, L.~W. Acton, C.~C. Kankelborg, C.~J. Wolfson, D.~J. Akin, M.~E.
  Bruner, R.~Caravalho, R.~C. Catura, R.~Chevalier, D.~W. Duncan, C.~G.
  Edwards, C.~N. Feinstein, S.~L. Freeland, F.~M. Friedlaender, C.~H. Hoffmann,
  N.~E. Hurlburt, B.~K. Jurcevich, N.~L. Katz, G.~A. Kelly, J.~R. Lemen,
  M.~Levay, R.~W. Lindgren, D.~P. Mathur, S.~B. Meyer, S.~J. Morrison, M.~D.
  Morrison, R.~W. Nightingale, T.~P. Pope, R.~A. Rehse, C.~J. Schrijver, R.~A.
  Shine, L.~Shing, K.~T. Strong, T.~D. Tarbell, A.~M. Title, D.~D. Torgerson,
  L.~Golub, J.~A. Bookbinder, D.~Caldwell, P.~N. Cheimets, W.~N. Davis, E.~E.
  Deluca, R.~A. McMullen, H.~P. Warren, D.~Amato, R.~Fisher, H.~Maldonado, and
  C.~Parkinson.
\newblock The transition region and coronal explorer.
\newblock {\em Sol. Phys.}, 187:229--260, 1999.

\bibitem{Ogawara_etal_1991}
Y.~Ogawara, T.~Takano, T.~Kato, T.~Kosugi, S.~Tsuneta, T.~Watanabe, I.~Kondo,
  and Y.~Uchida.
\newblock The solar-a mission - an overview.
\newblock {\em Sol. Phys.}, 136:1--16, 1991.

\bibitem{Pesnell_etal_2012}
W.~Dean Pesnell, B.~J. Thompson, and P.~C. Chamberlin.
\newblock The solar dynamics observatory ({SDO}).
\newblock {\em Sol. Phys.}, 275:3--15, 2012.

\bibitem{Kosugi_etal_2007}
T.~Kosugi, K.~Matsuzaki, T.~Sakao, T.~Shimizu, Y.~Sone, S.~Tachikawa,
  T.~Hashimoto, K.~Minesugi, A.~Ohnishi, T.~Yamada, S.~Tsuneta, H.~Hara,
  K.~Ichimoto, Y.~Suematsu, M.~Shimojo, T.~Watanabe, S.~Shimada, J.~M. Davis,
  L.~D. Hill, J.~K. Owens, A.~M. Title, J.~L. Culhane, L.~K. Harra, G.~A.
  Doschek, and L.~Golub.
\newblock The {Hinode} (solar-b) mission: An overview.
\newblock {\em Sol. Phys.}, 243:3--17, 2007.

\bibitem{Kaiser_etal_2008}
M.~L. Kaiser, T.~A. Kucera, J.~M. Davila, O.~C. {St. Cyr}, M.~Guhathakurta, and
  E.~Christian.
\newblock The stereo mission: An introduction.
\newblock {\em Space Sci. Rev.}, 136:5--16, 2008.

\bibitem{Muller_etal_2013}
D.~M{\"u}ller, R.~G. Marsden, O.~C. {St. Cyr}, and H.~R. Gilbert.
\newblock Solar orbiter . exploring the sun-heliosphere connection.
\newblock {\em Sol. Phys.}, 285:25--70, 2013.

\bibitem{Ogilvie_Parks_1996}
K.~W. Ogilvie and G.~K. Parks.
\newblock First results from {WIND} spacecraft: An introduction.
\newblock {\em Geophys. Res. Lett.}, 23:1179--1181, 1996.

\bibitem{Stone_etal_1998}
R.~G. Stone, A.~M. Frandsen, R.~A. Mewaldt, E.~R. Christian, D.~Margolies,
  J.~F. Ormes, and F.~Snow.
\newblock The advanced composition explorer.
\newblock {\em Space Sci. Rev.}, 86:1--22, 1998.

\bibitem{dscovr_2015}
NOAA.
\newblock Dscovr: Deep space climate observatory.
\newblock
  \url{https://www.nesdis.noaa.gov/content/dscovr-deep-space-climate-observatory},
  2015.

\bibitem{Winkler_1976}
W.~Winkler.
\newblock {HELIOS} assessment and mission results.
\newblock {\em Acta Astronautica}, 3:435--447, 1976.

\bibitem{Wenzel_etal_1992}
K.~P. Wenzel, R.~G. Marsden, D.~E. Page, and E.~J. Smith.
\newblock The {Ulysses} mission.
\newblock {\em Astron. \& Astrophys. Suppl.}, 92:207, 1992.

\bibitem{Fox_etal_2016}
N.~J. Fox, M.~C. Velli, S.~D. Bale, R.~Decker, A.~Driesman, R.~A. Howard, J.~C.
  Kasper, J.~Kinnison, M.~Kusterer, D.~Lario, M.~K. Lockwood, D.~J. McComas,
  N.~E. Raouafi, and A.~Szabo.
\newblock The solar probe plus mission: Humanity's first visit to our star.
\newblock {\em Space Sci. Rev.}, 204:7--48, 2016.

\bibitem{Solomon_etal_2007}
Sean~C. Solomon, Ralph~L. McNutt, Robert~E. Gold, and Deborah~L. Domingue.
\newblock {MESSENGER} mission overview.
\newblock {\em Space Sci. Rev.}, 131:3--39, 2007.

\bibitem{Svedhem_etal_2007}
H.~Svedhem, D.~V. Titov, D.~McCoy, J.-P. Lebreton, S.~Barabash, J.-L. Bertaux,
  P.~Drossart, V.~Formisano, B.~H{\"{a}}usler, O.~Korablev, W.~J. Markiewicz,
  D.~Nevejans, M.~P{\"{a}}tzold, G.~Piccioni, T.~L. Zhang, F.~W. Taylor,
  E.~Lellouch, D.~Koschny, O.~Witasse, H.~Eggel, M.~Warhaut, A.~Accomazzo,
  J.~Rodriguez-Canabal, J.~Fabrega, T.~Schirmann, A.~Clochet, and M.~Coradini.
\newblock {Venus Express} -- the first {European} mission to {Venus}.
\newblock {\em Planet. Space Sci.}, 55:1636--1652, 2007.

\bibitem{Schmidt_2003}
R.~Schmidt.
\newblock {Mars Express}-{ESA}'s first mission to planet {Mars}.
\newblock {\em Acta Astronautica}, 52:197--202, 2003.

\bibitem{Jakosky_etal_2015}
B.~M. Jakosky, R.~P. Lin, J.~M. Grebowsky, J.~G. Luhmann, D.~F. Mitchell,
  G.~Beutelschies, T.~Priser, M.~Acuna, L.~Andersson, D.~Baird, D.~Baker,
  R.~Bartlett, M.~Benna, S.~Bougher, D.~Brain, D.~Carson, S.~Cauffman,
  P.~Chamberlin, J.~Y. Chaufray, O.~Cheatom, J.~Clarke, J.~Connerney,
  T.~Cravens, D.~Curtis, G.~Delory, S.~Demcak, A.~DeWolfe, F.~Eparvier,
  R.~Ergun, A.~Eriksson, J.~Espley, X.~Fang, D.~Folta, J.~Fox, C.~Gomez-Rosa,
  S.~Habenicht, J.~Halekas, G.~Holsclaw, M.~Houghton, R.~Howard, M.~Jarosz,
  N.~Jedrich, M.~Johnson, W.~Kasprzak, M.~Kelley, T.~King, M.~Lankton,
  D.~Larson, F.~Leblanc, F.~Lefevre, R.~Lillis, P.~Mahaffy, C.~Mazelle,
  W.~McClintock, J.~McFadden, D.~L. Mitchell, F.~Montmessin, J.~Morrissey,
  W.~Peterson, W.~Possel, J.~A. Sauvaud, N.~Schneider, W.~Sidney, S.~Sparacino,
  A.~I.~F. Stewart, R.~Tolson, D.~Toublanc, C.~Waters, T.~Woods, R.~Yelle, and
  R.~Zurek.
\newblock The mars atmosphere and volatile evolution ({MAVEN}) mission.
\newblock {\em Space Sci. Rev.}, 195:3--48, 2015.

\bibitem{WangY_etal_2020}
Yamin Wang, Xin Chen, Pengcheng Wang, Chengbo Qiu, Yuming Wang, and Yonghe
  Zhang.
\newblock Concept of the solar ring mission: Preliminary design and mission
  profile.
\newblock {\em Sci. China Tech. Sci.}, submitted, 2020.

\bibitem{Gary_Hagyard_1990}
G.~Allen Gary and M.~J. Hagyard.
\newblock Transformation of vector magnetograms and the problems associated
  with the effects of perspective and the azimuthal ambiguity.
\newblock {\em Sol. Phys.}, 126:21--36, 1990.

\bibitem{Schou_etal_2012}
J.~Schou, P.~H. Scherrer, R.~I. Bush, R.~Wachter, S.~Couvidat, M.~C.
  Rabello-Soares, R.~S. Bogart, J.~T. Hoeksema, Y.~Liu, T.~L. Duvall, D.~J.
  Akin, B.~A. Allard, J.~W. Miles, R.~Rairden, R.~A. Shine, T.~D. Tarbell,
  A.~M. Title, C.~J. Wolfson, D.~F. Elmore, A.~A. Norton, and S.~Tomczyk.
\newblock Design and ground calibration of the helioseismic and magnetic imager
  ({HMI}) instrument on the solar dynamics observatory ({SDO}).
\newblock {\em Sol. Phys.}, 275:229--259, 2012.

\bibitem{LiuL_etal_2016}
Lijuan Liu, Yuming Wang, Jingxiu Wang, Chenglong Shen, Pinzhong Ye, Rui Liu,
  Jun Chen, Quanhao Zhang, and S.~Wang.
\newblock Why is a flare-rich active region {CME}-poor?
\newblock {\em Astrophys. J.}, 826:119(10pp), 2016.

\bibitem{JinC_etal_2012}
C.~L. Jin, J.~X. Wang, and Z.~X. Xie.
\newblock Solar intranetwork magnetic elements: Intrinsically weak or strong?
\newblock {\em Sol. Phys.}, 280:51--67, 2012.

\bibitem{Wiegelmann_Sakurai_2012}
Thomas Wiegelmann and Takashi Sakurai.
\newblock Solar force-free magnetic fields.
\newblock {\em Living Rev. Sol. Phys.}, 9:5, 2012.

\bibitem{Wiegelmann_2008}
Thomas Wiegelmann.
\newblock Nonlinear force-free modeling of the solar coronal magnetic field.
\newblock {\em J. Geophys. Res.: Space Phys.}, 113:A03S02, 2008.

\bibitem{WangY_etal_2020a}
Ya~Wang, Yingna Su, Yuming Wang, Yuanyong Deng, Rui Liu, Jianping Li, and
  Haisheng Ji.
\newblock Mapping global vector magnetic field of the {Sun}'s photosphere
  without 180-degree ambiguity.
\newblock {\em Sci. China Phys. Mech. \& Astron.}, in preparatoin, 2020.

\bibitem{Schrijver_Title_2011}
C.~J. Schrijver and A.~M. Title.
\newblock Long-range magnetic couplings between solar flares and coronal mass
  ejections observed by {SDO} and {STEREO}.
\newblock {\em J. Geophys. Res.: Space Phys.}, 116:A04108, 2011.

\bibitem{Christensen-Dalsgaard_etal_1996}
J.~Christensen-Dalsgaard, W.~Dappen, S.~V. Ajukov, E.~R. Anderson, H.~M. Antia,
  S.~Basu, V.~A. Baturin, G.~Berthomieu, B.~Chaboyer, S.~M. Chitre, A.~N. Cox,
  P.~Demarque, J.~Donatowicz, W.~A. Dziembowski, M.~Gabriel, D.~O. Gough, D.~B.
  Guenther, J.~A. Guzik, J.~W. Harvey, F.~Hill, G.~Houdek, C.~A. Iglesias,
  A.~G. Kosovichev, J.~W. Leibacher, P.~Morel, C.~R. Proffitt, J.~Provost,
  J.~Reiter, E.~J. {Jr. Rhodes}, F.~J. Rogers, I.~W. Roxburgh, M.~J. Thompson,
  and R.~K. Ulrich.
\newblock The current state of solar modeling.
\newblock {\em Science}, 272:1286--1292, 1996.

\bibitem{Scherrer_etal_1995}
P.~H. Scherrer, R.~S. Bogart, R.~I. Bush, J.~T. Hoeksema, A.~G. Kosovichev,
  J.~Schou, W.~Rosenberg, L.~Springer, T.~D. Tarbell, A.~Title, C.~J. Wolfson,
  I.~Zayer, and {MDI Engineering Team}.
\newblock The solar oscillations investigation - {Michelson Doppler Imager}.
\newblock {\em Sol. Phys.}, 162:129--188, 1995.

\bibitem{Harvey_etal_1996}
J.~W. Harvey, F.~Hill, R.~P. Hubbard, J.~R. Kennedy, J.~W. Leibacher, J.~A.
  Pintar, P.~A. Gilman, R.~W. Noyes, A.~M. Title, J.~Toomre, R.~K. Ulrich,
  A.~Bhatnagar, J.~A. Kennewell, W.~Marquette, J.~Patron, O.~Saa, and
  E.~Yasukawa.
\newblock The global oscillation network group ({GONG}) project.
\newblock {\em Science}, 272:1284--1286, 1996.

\bibitem{Thompson_etal_1996}
M.~J. Thompson, J.~Toomre, E.~R. Anderson, H.~M. Antia, G.~Berthomieu,
  D.~Burtonclay, S.~M. Chitre, J.~Christensen-Dalsgaard, T.~Corbard, M.~{De
  Rosa}, C.~R. Genovese, D.~O. Gough, D.~A. Haber, J.~W. Harvey, F.~Hill,
  R.~Howe, S.~G. Korzennik, A.~G. Kosovichev, J.~W. Leibacher, F.~P. Pijpers,
  J.~Provost, E.~J. {Jr. Rhodes}, J.~Schou, T.~Sekii, P.~B. Stark, and P.~R.
  Wilson.
\newblock Differential rotation and dynamics of the solar interior.
\newblock {\em Science}, 272:1300--1305, 1996.

\bibitem{Howe_etal_2000}
R.~Howe, J.~Christensen-Dalsgaard, F.~Hill, R.~W. Komm, R.~M. Larsen, J.~Schou,
  M.~J. Thompson, and J.~Toomre.
\newblock Deeply penetrating banded zonal flows in the solar convection zone.
\newblock {\em Astrophys. J. Lett.}, 533:L163--L166, 2000.

\bibitem{ZhaoJ_etal_2013}
Junwei Zhao, R.~S. Bogart, A.~G. Kosovichev, T.~L. {Jr. Duvall}, and Thomas
  Hartlep.
\newblock Detection of equatorward meridional flow and evidence of double-cell
  meridional circulation inside the {Sun}.
\newblock {\em Astrophys. J. Lett.}, 774:L29(6pp), 2013.

\bibitem{Miesch_Brown_2012}
Mark~S. Miesch and Benjamin~P. Brown.
\newblock Convective {Babcock-Leighton} dynamo models.
\newblock {\em Astrophys. J. Lett.}, 746:L26(5pp), 2012.

\bibitem{Simnett_Hudson_1997}
G.~M. Simnett and H.~S. Hudson.
\newblock The evolution of a rapidly-expanding active region loop into a
  trans-equatorial coronal mass ejection.
\newblock In {\em Correlated Phenomena at the Sun, in the Heliosphere and in
  Geospace}, Proc. 31st ESLAB Symp. (ESA SP-415), pages 437--441, The
  Netherlands, 1997.

\bibitem{Moon_etal_2003}
Y.-J. Moon, G.~S. Choe, Haimin Wang, and Y.~D. Park.
\newblock Sympathetic coronal mass ejections.
\newblock {\em Astrophys. J.}, 588:1176--1182, 2003.

\bibitem{ZhouG_etal_2007}
Guiping Zhou, Jingxiu Wang, Yuming Wang, and Yuzong Zhang.
\newblock Quasi-simultaneous flux emergence in the events of {October/November}
  2003.
\newblock {\em Sol. Phys.}, 244:13--24, 2007.

\bibitem{ZhangY_etal_2007}
Yuzong Zhang, Jingxiu Wang, Gemma D.~R. Attrill, Louise~K. Harra, Zhiliang
  Yang, and Xiangtao He.
\newblock Coronal magnetic connectivity and {EUV} dimmings.
\newblock {\em Sol. Phys.}, 241:329--349, 2007.

\bibitem{Pevtsov_2000}
A.~A. Pevtsov.
\newblock Transequatorial loops in the solar corona.
\newblock {\em Astrophys. J.}, 531:553--560, 2000.

\bibitem{Heinemann_etal_2016}
Stephan~G. Heinemann, Manuela Temmer, Stefan~J. Hofmeister, Astrid~M. Veronig,
  and Susanne Vennerstrom.
\newblock Three-phase evolution of a coronal hole. {I.} $360^\circ$ remote
  sensing and in situ observations.
\newblock {\em Astrophys. J.}, 861:151(12pp), 2016.

\bibitem{LiuY_etal_2012a}
Y.~Liu, J.~T. Hoeksema, P.~H. Scherrer, J.~Schou, S.~Couvidat, R.~I. Bush,
  T.~L. Duvall, K.~Hayashi, X.~Sun, and X.~Zhao.
\newblock Comparison of line-of-sight magnetograms taken by the solar dynamics
  observatory/helioseismic and magnetic imager and solar and heliospheric
  observatory/michelson doppler imager.
\newblock {\em Sol. Phys.}, 279:295--316, 2012.

\bibitem{LiX_etal_2020}
Xiaolei Li, Yuming Wang, Rui Liu, Chenglong Shen, Quanhao Zhang, Shaoyu Lv, Bin
  Zhuang, Fang Shen, Jiajia Liu, and Yutian Chi.
\newblock Reconstructing solar wind inhomogeneous structures from stereoscopic
  observations in white-light: Solar wind transients in 3d.
\newblock {\em J. Geophys. Res.: Space Phys.}, submitted, 2020.

\bibitem{Aschwanden_etal_2008}
Markus~J. Aschwanden, Jean-Pierre Wuelser, Nariaki~V. Nitta, and James~R.
  Lemen.
\newblock First three-dimensional reconstructions of coronal loops with the
  {STEREO} {A} and {B} spacecraft. {I}. geometry.
\newblock {\em Astrophys. J.}, 679:827--842, 2008.

\bibitem{LiuJ_etal_2014}
Jiajia Liu, Yuming Wang, Rui Liu, Quanhao Zhang, Kai Liu, Chenglong Shen, and
  S.~Wang.
\newblock When and how does a prominence-like jet gain kinetic energy?
\newblock {\em Astrophys. J.}, 782:94(7pp), 2014.

\bibitem{Kwon_etal_2010}
Ryun-Young Kwon, Jongchul Chae, and Jie Zhang.
\newblock Stereoscopic determination of heights of extreme ultraviolet bright
  points using data taken by {SECCHI/EUVI} aboard {STEREO}.
\newblock {\em Astrophys. J.}, 714:130--137, 2010.

\bibitem{Robbrecht_etal_2009}
Eva Robbrecht, Spiros Patsourakos, and Angelos Vourlidas.
\newblock No trace left behind: {STEREO} observation of a coronal mass ejection
  without low coronal signatures.
\newblock {\em Astrophys. J.}, 701:283--291, 2009.

\bibitem{Wang_etal_2011}
Yuming Wang, Caixia Chen, Bin Gui, Chenglong Shen, Pinzhong Ye, and S.~Wang.
\newblock Statistical study of coronal mass ejection source locations:
  Understanding cmes viewed in coronagraphs.
\newblock {\em J. Geophys. Res.: Space Phys.}, 116:A04104,
  doi:{10.1029/2010JA016101}, 2011.

\bibitem{Thernisien_etal_2006}
A.~Thernisien, R.A. Howard, and A.~Vourlidas.
\newblock Modeling of flux rope coronal mass ejections.
\newblock {\em Astrophys. J.}, 652:763--773, 2006.

\bibitem{Sheeley_etal_2009}
N.~R. {Sheeley, Jr.}, D.~D.-H. Lee, K.~P. Casto, Y.-M. Wang, and N.~B. Rich.
\newblock The structure of streamer blobs.
\newblock {\em Astrophys. J.}, 694:1471--1480, 2009.

\bibitem{Lugaz_etal_2009}
N.~Lugaz, A.~Vourlidas, and I.~I. Roussev.
\newblock Deriving the radial distances of wide coronal mass ejections from
  elongation measurements in the heliosphere application to {CME-CME}
  interaction.
\newblock {\em Ann. Geophys.}, 27:3479--3488, 2009.

\bibitem{FengL_etal_2013}
L.~Feng, B.~Inhester, and M.~Mierla.
\newblock Comparisons of {CME} morphological characteristics derived from five
  {3D} reconstruction methods.
\newblock {\em Sol. Phys.}, 282:221--238, 2013.

\bibitem{LiX_etal_2018}
Xiaolei Li, Yuming Wang, Rui Liu, Chenglong Shen, Quanhao Zhang, Bin Zhuang,
  Jiajia Liu, and Yutian Chi.
\newblock Reconstructing solar wind inhomogeneous structures from stereoscopic
  observations in white-light: Small transients along the {Sun-Earth} line.
\newblock {\em J. Geophys. Res.: Space Phys.}, 123:7257--7270, 2018.

\bibitem{Wang_etal_2016a}
Yuming Wang, Quanhao Zhang, Jiajia Liu, Chenglong Shen, Fang Shen, Zicai Yang,
  T.~Zic, B.~Vrsnak, D.~F. Webb, Rui Liu, S.~Wang, Jie Zhang, Qiang Hu, and Bin
  Zhuang.
\newblock On the propagation of a geoeffective coronal mass ejection during
  15-–17 {March} 2015.
\newblock {\em J. Geophys. Res.: Space Phys.}, 121:7423–--7434, 2016.

\bibitem{Lyu_etal_2020}
Shaoyu Lyu, Xiaolei Li, and Yuming Wang.
\newblock Optimal stereoscopic angle for reconstructing solar wind
  inhomogeneous structures.
\newblock {\em Sci. China Phys. Mech. \& Astron.}, submitted, 2020.

\bibitem{Wang_etal_2004b}
Yuming Wang, Chenglong Shen, Pinzhong Ye, and S.~Wang.
\newblock Deflection of coronal mass ejection in the interplanetary medium.
\newblock {\em Sol. Phys.}, 222:329--343, 2004.

\bibitem{Riley_Crooker_2004}
Pete Riley and N.~U. Crooker.
\newblock Kinematic treatment of coronal mass ejection evolution in the solar
  wind.
\newblock {\em Astrophys. J.}, 600:1035--1042, 2004.

\bibitem{Manchester_etal_2004}
IV~Manchester, W., T.~Gombosi, D.~DeZeeuw, and Y.~Fan.
\newblock Eruption of a buoyantly emerging magnetic flux rope.
\newblock {\em Astrophys. J.}, 610:588--596, 2004.

\bibitem{Wang_etal_2014}
Yuming Wang, Boyi Wang, Chenglong Shen, Fang Shen, and No\'{e} Lugaz.
\newblock Deflected propagation of a coronal mass ejection from the corona to
  interplanetary space.
\newblock {\em J. Geophys. Res.: Space Phys.}, 119:5117--5132, 2014.

\bibitem{Kay_Opher_2015}
C.~Kay and M.~Opher.
\newblock The heliocentric distance where the deflections and rotations of
  solar coronal mass ejections occur.
\newblock {\em Astrophys. J. Lett.}, 811:L36(6pp), 2015.

\bibitem{Gopalswamy_etal_2000}
N.~Gopalswamy, A.~Lara, R.~P. Lepping, M.~L. Kaiser, D.~Berdichevsky, and O.~C.
  {St. Cyr}.
\newblock Interplanetary acceleration of coronal mass ejections.
\newblock {\em Geophys. Res. Lett.}, 27:145--148, 2000.

\bibitem{Vrsnak_etal_2008}
B.~Vr\v{s}nak, D.~Vrbanec, and J.~\v{C}alogovi\'c.
\newblock Dynamics of coronal mass ejections. the mass-scaling of the
  aerodynamic drag.
\newblock {\em Astron. \& Astrophys.}, 490:811--815, 2008.

\bibitem{Vrsnak_etal_2013}
B.~Vr\v{s}nak, T.~\v{Z}ic, D.~Vrbanec, M.~Temmer, T.~Rollett, C.~M{\"o}stl,
  A.~Veronig, J.~\v{C}alogovi\'c, M.~Dumbovi\'c, S.~Luli\'c, Y.-J. Moon, and
  A.~Shanmugaraju.
\newblock Propagation of interplanetary coronal mass ejections: The drag-based
  model.
\newblock {\em Sol. Phys.}, 285:295--315, 2013.

\bibitem{Shen_etal_2014}
Chenglong Shen, Yuming Wang, Zonghao Pan, Bin Miao, Pinzhong Ye, and S.~Wang.
\newblock Full-halo coronal mass ejections: Arrival at the {Earth}.
\newblock {\em J. Geophys. Res.: Space Phys.}, 119:5107--5116, 2014.

\bibitem{Dasso_etal_2006}
S.~Dasso, C.~H. Mandrini, P.~D\'{e}moulin, and M.~L. Luoni.
\newblock A new model-independent method to compute magnetic helicity in
  magnetic clouds.
\newblock {\em Astron. \& Astrophys.}, 455:349--359, 2006.

\bibitem{Ruffenach_etal_2015}
A.~Ruffenach, B.~Lavraud, C.~J. Farrugia, P.~Demoulin, S.~Dasso, M.~J. Owens,
  J.-A. Sauvaud, A.~P. Rouillard, A.~Lynnyk, C.~Foullon, N.~P. Savani, J.~G.
  Luhmann, and A.~B. Galvin.
\newblock Statistical study of magnetic cloud erosion by magnetic reconnection.
\newblock {\em J. Geophys. Res.: Space Phys.}, 120:43--60, 2015.

\bibitem{Wang_etal_2018}
Yuming Wang, Chenglong Shen, Rui Liu, Mengjiao Xu, Qiang Hu, Jiajia Liu,
  Jingnan Guo, Xiaolei Li, , and Tielong Zhang.
\newblock Understanding the twist distribution inside magnetic flux ropes by
  anatomizing an interplanetary magnetic cloud.
\newblock {\em J. Geophys. Res.: Space Phys.}, 123:3238–--3261, 2018.

\bibitem{Shen_etal_2012}
Chenglong Shen, Yuming Wang, Shui Wang, Ying Liu, Rui Liu, Angelos Vourlidas,
  Bin Miao, Pinzhong Ye, Jiajia Liu, and Zhenjun Zhou.
\newblock Super-elastic collision of large-scale magnetized plasmoids in the
  heliosphere.
\newblock {\em Nature Phys.}, 8:923--928, 2012.

\bibitem{Lugaz_etal_2012}
N.~Lugaz, C.~J. Farrugia, J.~A. Davies, C.~Mostl, C.~J. Davis, I.~I. Roussev,
  and M.~Temmer.
\newblock The deflection of the two interacting coronal mass ejections of 2010
  may 23-24 as revealed by combined in site measurements and heliospheric
  imaging.
\newblock {\em Astrophys. J.}, 759:68(13pp), 2012.

\bibitem{Temmer_etal_2014}
Manuela Temmer, A.~M. Veronig, V.~Peinhart, and Bojan Vr\v{s}nak.
\newblock Asymmetry in the {CME-CME} interaction process for the events from
  2011 {February} 14--15.
\newblock {\em Astrophys. J.}, 785:85(7pp), 2014.

\bibitem{Mishra_etal_2017}
Wageesh Mishra, Yuming Wang, Nandita Srivastava, and Chenglong Shen.
\newblock Assessing the nature of collisions of coronal mass ejections in the
  inner heliosphere.
\newblock {\em Astrophys. J. Suppl. Ser.}, 232:5(24pp), 2017.

\bibitem{Larson_etal_1997}
D.~E. Larson, R.~P. Lin, J.~M. McTiernan, J.~P. McFadden, R.~E. Ergun,
  M.~McCarthy, H.~R\`{e}me, T.~R. Sanderson, M.~Kaiser, R.~P. Lepping, and
  J.~Mazur.
\newblock Tracing the topology of the october 18-20, 1995, magnetic cloud with
  $\sim0.1-10^2$ kev electrons.
\newblock {\em Geophys. Res. Lett.}, 24(15):1911--1914, 1997.

\bibitem{Wang_etal_2015}
Yuming Wang, Zhenjun Zhou, Chenglong Shen, Rui Liu, and S.~Wang.
\newblock Investigating plasma motion of magnetic clouds at 1 {AU} through a
  velocity-modified cylindrical force-free flux rope model.
\newblock {\em J. Geophys. Res.: Space Phys.}, 120:1543--1565, 2015.

\bibitem{Hassler_etal_2012}
D.~M. Hassler, C.~Zeitlin, R.~F. Wimmer-Schweingruber, S.~B{\"{o}}ttcher,
  C.~Martin, J.~Andrews, E.~B{\"{o}}hm, D.~E. Brinza, M.~A. Bullock,
  S.~Burmeister, B.~Ehresmann, M.~Epperly, D.~Grinspoon, J.~K{\"{o}}hler,
  O.~Kortmann, K.~Neal, J.~Peterson, A.~Posner, S.~Rafkin, L.~Seimetz, K.~D.
  Smith, Y.~Tyler, G.~Weigle, G.~Reitz, and F.~A. Cucinotta.
\newblock The {Radiation Assessment Detector} ({RAD}) investigation.
\newblock {\em Space Sci. Rev.}, 170:503--558, 2012.

\bibitem{GuoJ_etal_2018a}
Jingnan Guo, Mateja Dumbovi\'c, Robert~F. Wimmer-Schweingruber, Manuela Temmer,
  Henning Lohf, Yuming Wang, Astrid Veronig, Donald~M. Hassler, Leila~M. Mays,
  Cary Zeitlin, Bent Ehresmann, Olivier Witasse, Johan~L. {Freiherr von
  Forstner}, Bernd Heber, Mats Holmstr{\"o}m, and Arik Posner.
\newblock Modeling the evolution and propagation of 10 {September} 2017 {CMEs}
  and {SEPs} arriving at {Mars} constrained by remote sensing and in situ
  measurement.
\newblock {\em Space Weather}, 16:1156--1169, 2018.

\bibitem{Wang_etal_2016}
Yuming Wang, Bin Zhuang, Qiang Hu, Rui Liu, Chenglong Shen, and Yutian Chi.
\newblock On the twists of interplanetary magnetic flux ropes observed at 1
  {AU}.
\newblock {\em J. Geophys. Res.: Space Phys.}, 121:9316--9339, 2016.

\bibitem{Demoulin_etal_2016}
P.~D\'{e}moulin, M.~Janvier, and S.~Dasso.
\newblock Magnetic flux and helicity of magnetic clouds.
\newblock {\em Sol. Phys.}, 291:531--557, 2016.

\bibitem{Owens_2016}
M.~J. Owens.
\newblock Do the legs of magnetic clouds contain twisted flux-rope magnetic
  fields?
\newblock {\em Astrophys. J.}, 818:197(5pp), 2016.

\bibitem{ZhaoA_etal_2017}
Ake Zhao, Yuming Wang, Yutian Chi, Jiajia Liu, Chenglong Shen, and Rui Liu.
\newblock Main cause of the poloidal plasma motion inside a magnetic cloud
  inferred from multiple-spacecraft observations.
\newblock {\em Sol. Phys.}, 292:58, 2017.

\bibitem{Owens_etal_2017}
M.~J. Owens, M.~Lockwood, and L.~A. Barnard.
\newblock Coronal mass ejections are not coherent magnetohydrodynamic
  structures.
\newblock {\em Sci. Rep.}, 7:4152, 2017.

\bibitem{Desai_Giacalone_2016}
Mihir Desai and Joe Giacalone.
\newblock Large gradual solar energetic particle events.
\newblock {\em Living Rev. Sol. Phys.}, 13:3(132pp), 2016.

\bibitem{Cane_etal_1988}
H.~V. Cane, D.~V. Reames, and T.~T {von Rosenvinge}.
\newblock The role of interplanetary shocks in the longitude distribution of
  solar energetic particles.
\newblock {\em J. Geophys. Res.: Space Phys.}, 93:9555--9567, 1988.

\bibitem{Wang_etal_2009}
Yuming Wang, Jie Zhang, and Chenglong Shen.
\newblock An analytical model probing the internal state of coronal mass
  ejections based on observations of their expansions and propagations.
\newblock {\em J. Geophys. Res.: Space Phys.}, 114:A10104, 2009.

\bibitem{Mishra_Wang_2018}
Wageesh Mishra and Yuming Wang.
\newblock Modeling the thermodynamic evolution of coronal mass ejections using
  their kinematics.
\newblock {\em Astrophys. J.}, 865:50(15pp), 2018.

\bibitem{Wang_etal_2010}
Yuming Wang, Hao Cao, Junhong Chen, Tengfei Zhang, Sijie Yu, Huinan Zheng,
  Chenglong Shen, Jie Zhang, and S.~Wang.
\newblock Solar limb prominence catcher and tracker ({SLIPCAT}): An automated
  system and its preliminary statistical results.
\newblock {\em Astrophys. J.}, 717:973--986, 2010.

\bibitem{Gosling_2012}
J.~T. Gosling.
\newblock Magnetic reconnection in the solar wind.
\newblock {\em Space Sci. Rev.}, 172:187--200, 2012.

\bibitem{ShenF_etal_2013}
Fang Shen, Chenglong Shen, Yuming Wang, Xueshang Feng, and Changqing Xiang.
\newblock Could the collision of cmes in the heliosphere be super-elastic?
  validation through three-dimensional simulations.
\newblock {\em Geophys. Res. Lett.}, 40:1457--1461, 2013.

\bibitem{Reiner_Stone_1986}
M.~J. Reiner and R.~G. Stone.
\newblock A new method for reconstructing type-{III} trajectories.
\newblock {\em Sol. Phys.}, 106:397--401, 1986.

\bibitem{Krupar_etal_2014}
V.~Krupar, M.~Maksimovic, O.~Santolik, B.~Cecconi, and O.~Kruparova.
\newblock Statistical survey of type {III} radio bursts at long wavelengths
  observed by the solar terrestrial relations observatory ({STEREO})/waves
  instruments: Goniopolarimetric properties and radio source locations.
\newblock {\em Sol. Phys.}, 289:4633--4652, 2014.

\bibitem{Magdalenic_etal_2014}
J.~Magdaleni{\'c}, C.~Marqu{\'e}, V.~Krupar, M.~Mierla, A.~N. Zhukov,
  L.~Rodriguez, M.~Maksimovi{\'c}, and B.~Cecconi.
\newblock Tracking the {CME}-driven shock wave on 2012 {March} 5 and radio
  triangulation of associated radio emission.
\newblock {\em Astrophys. J.}, 791:115(14pp), 2014.

\bibitem{ZhangP_etal_2019}
Peijin Zhang, Chuangbin Wang, Ye~Lin, and Yuming Wang.
\newblock Forward modeling of the type {III} radio burst exciter.
\newblock {\em Sol. Phys.}, 294:62(20pp), 2019.

\bibitem{Cecconi_etal_2008}
B.~Cecconi, X.~Bonnin, S.~Hoang, M.~Maksimovic, S.~D. Bale, J.-L. Bougeret,
  K.~Goetz, A.~Lecacheux, M.~J. Reiner, H.~O. Rucker, and P.~Zarka.
\newblock {STEREO}/waves goniopolarimetry.
\newblock {\em Space Sci. Rev.}, 136:549--563, 2008.

\bibitem{Leamon_etal_1998}
Robert~J. Leamon, Charles~W. Smith, Norman~F. Ness, William~H. Matthaeus, and
  Hung~K. Wong.
\newblock Observational constraints on the dynamics of the interplanetary
  magnetic field dissipation range.
\newblock {\em 103}, 103:4775--4788, 1998.

\bibitem{HuR_etal_2019}
Renxiang Hu, Xu~Shan, Guangyuan Yuan, Shuwen Wang, Weihang Zhang, Wei Qi, Zhe
  Cao, Yiren Li, Manming Chen, Xiaoping Yang, Bo~Wang, Sipei Shao, Xinjun Hao,
  Changqing Feng, Zhenpeng Su, Chenglong Shen, Li~Xin, Dai Guyue, Binglin Qiu,
  Zonghao Pan, Kai Liu, Chunkai Xu, Shubin Liu, An~Qi, Tielong Zhang, Yuming
  Wang, and Xiangjun Chen.
\newblock A low-energy ion spectrometer with half-space entrance for three-axis
  stabilized spacecraft.
\newblock {\em Sci. China Tech. Sci.}, 62:1015--1027, 2019.

\bibitem{Malandraki_etal_2005}
O.~E. Malandraki, D.~Lario, L.~J. Lanzerotti, E.~T. Sarris, A.~Geranios, and
  G.~Tsiropoula.
\newblock {October/November} 2003 interplanetary coronal mass ejections:
  {ACE/EPAM} solar energetic particle observations.
\newblock {\em J. Geophys. Res.: Space Phys.}, 110:A09S06, 2005.

\bibitem{WuW_etal_2018}
Weiren Wu, Ming Chen, Zhe Zhang, Xiangnan Liu, and Yuhui Dong.
\newblock Overview of deep space laser communication.
\newblock {\em Sci. China Inf. Sci.}, 61:040301, 2018.

\end{thebibliography}

\end{multicols}

\end{document}